\documentclass{article}

\usepackage[title]{appendix}

\usepackage{pgfplots}
\pgfplotsset{compat=1.15}
\usepackage{mathrsfs}
\usetikzlibrary{arrows}

\usepackage{widetext}

\usepackage{wrapfig}
\usepackage[font=footnotesize,labelfont=bf]{caption}

\usepackage[utf8]{inputenc}
\usepackage[T1]{fontenc}
\usepackage{dsfont}
\usepackage{amsfonts}
\usepackage{tgtermes}

\usepackage{xcolor,colortbl}

\usepackage{enumitem}
\setlist[itemize]{itemsep=0mm}

\usepackage{graphicx}
\graphicspath{{images/}}
\usepackage[a4paper,width=150mm,top=25mm,bottom=25mm,left=25mm,right=15mm]{geometry}

\usepackage[T1]{fontenc}
\usepackage{xcolor}
\definecolor{RoyalRed}{RGB}{157,16, 45}
\usepackage{titlesec}
\titleformat{\chapter}[display]
  {\normalsize \huge  \color{black}}%
  {\flushright\normalsize \color{RoyalRed}%
   \sc{\chaptertitlename}\hspace{1ex}%
   {\fontsize{60}{60}\selectfont\thechapter}
   \noindent\makebox[\linewidth]{\rule{\paperwidth}{0.4pt}}}%
  {10 pt}%
  {\bfseries\huge}%

\usepackage{sidecap}

\usepackage{makecell}

\def\T{{ \mathrm{\scriptscriptstyle T} }}
\def\pvec{\boldsymbol{\theta}}
\def\hpvec{\boldsymbol{\hat{\theta}}}

\def\d{\mathrm{d}}

\def\df{\mathbf{d}_\f}
\def\ds{\mathbf{d}_\s}

\def\Sf{\mathbf{s}_\f}

\def\Sn{S_\mathrm{n}}
\def\fish{\mathbf{F}}
\def\f{\mathrm{f}}
\def\s{\mathrm{s}}
\def\Df{\mathcal{D}_\f}
\def\Ds{\mathcal{D}_\s}

\def\Nf{N_\f}
\def\Ns{N_\s}
\def\cvm{\boldsymbol{\Sigma}}
\def\icvm{\boldsymbol{\Sigma}^{-1}}
\def\xcvm{\boldsymbol{\Gamma}}
\def\xicvm{\boldsymbol{\Gamma}^{-1}}
\usepackage{amsmath,mleftright}
\usepackage{xparse}
\usepackage{mathtools}
\NewDocumentCommand{\evalat}{sO{\Bigg}mm}{%
  \IfBooleanTF{#1}
   {\mleft. #3 \mright|_{#4}}
   {#3#2|_{#4}}%
}

\DeclareMathAlphabet{\mathpzc}{OT1}{pzc}{m}{it}

\usepackage[breakable, theorems, skins]{tcolorbox}
\tcbset{enhanced}

\usepackage{amsmath}
\usepackage{units}
\usepackage{mathtools}

\usepackage{soulutf8}

\setul{0.2ex}{0.8pt}

\usepackage{bm}

\usepackage{cancel}
\usepackage{color}
\usepackage{caption}
\usepackage{array}
\usepackage{changepage}
\usepackage{verbatimbox}
\usepackage{multicol}
\usepackage{subcaption}
\usepackage[colorlinks,allcolors=blue]{hyperref}

\usepackage[backend=biber,style=ieee,citestyle=numeric-comp,sorting=none]{biblatex} 
\usepackage{amsthm}
\theoremstyle{definition}
\newtheorem*{definition*}{Definition}

\renewcommand{\arraystretch}{1.0}

\makeatletter
\renewcommand\paragraph{\@startsection{paragraph}{4}{\z@}%
            {-2.5ex\@plus -1ex \@minus -.25ex}%
            {1.25ex \@plus .25ex}%
            {\normalfont\normalsize\bfseries}}
\makeatother
\def\T{{ \mathrm{\scriptscriptstyle T} }}
\def\pvec{\boldsymbol{\theta}}
\def\hpvec{\boldsymbol{\hat{\theta}}}

\def\d{\mathrm{d}}

\def\df{\mathbf{d}_\f}
\def\ds{\mathbf{d}_\s}

\def\Sf{\mathbf{s}_\f}

\def\Sn{S_\mathrm{n}}
\def\fish{\mathbf{F}}
\def\f{\mathrm{f}}
\def\s{\mathrm{s}}
\def\Df{\mathcal{D}_\f}
\def\Ds{\mathcal{D}_\s}

\def\Nf{N_\f}
\def\Ns{N_\s}
\def\cvm{\mathbf{C}}
\def\icvm{\mathbf{C}^{-1}}
\def\xcvm{\boldsymbol{\Gamma}}
\def\xicvm{\boldsymbol{\Gamma}^{-1}}

\usepackage{soul}

\usepackage{amsmath,mleftright}
\usepackage{xparse}
\usepackage{mathtools}

\usepackage{fancyvrb}

\VerbatimFootnotes

\DeclarePairedDelimiter\abs{\lvert}{\rvert}%
\DeclarePairedDelimiter\norm{\lVert}{\rVert}%

\DeclareMathOperator\supp{supp}

\makeatletter
\let\oldabs\abs
\def\abs{\@ifstar{\oldabs}{\oldabs*}}
\let\oldnorm\norm
\def\norm{\@ifstar{\oldnorm}{\oldnorm*}}
\makeatother

\makeatletter
\newenvironment{tablehere}
  {\def\@captype{table}}
  {}

\newenvironment{figurehere}
  {\def\@captype{figure}}
  {}
\makeatother

\usepackage{cases}
\usepackage{graphicx}
\usepackage{verbatim}
\usepackage{amssymb}
\newcommand\numberthis{\addtocounter{equation}{1}\tag{\theequation}}

\definecolor{ttqqqq}{rgb}{0.2,0,0}

\usepackage{authblk}

\usepackage[acronym,nomain,nonumberlist]{glossaries}
\makeglossaries

\newacronym{acf}{ACF}{autocorrelation function}
\newacronym{afr}{AFR}{adaptive frequency resolution}
\newacronym{asd}{ASD}{amplitude spectral density}
\newacronym{bh}{BH}{black hole}
\newacronym[longplural={black hole binaries}]{bhb}{BHB}{black hole binary}

\newacronym{cbc}{CBC}{compact binary coalescence}
\newacronym{ce}{CE}{Cosmic Explorer}
\newacronym{cmb}{CMB}{cosmic microwave background}
\newacronym{cmjs}{CMJS}{combined marginal Jenson-Shannon}
\newacronym{cw}{CW}{continuous wave}
\newacronym{dcs}{dCS}{dynamical Chern-Simons}
\newacronym{decigo}{DECIGO}{DECi-hertz Interferometer Gravitational wave Observatory}
\newacronym{dft}{DFT}{discrete Fourier transform}
\newacronym{dof}{DOF}{degree of freedom}
\newacronym{edgb}{EdGB}{Einstein-dilaton Gauss-Bonnett}
\newacronym{efes}{EFEs}{Einstein field equations}
\newacronym{et}{ET}{Einstein Telescope}
\newacronym{fft}{FFT}{fast Fourier transform}
\newacronym{fim}{FIM}{Fisher information matrix}
\newacronym{gb}{GB}{Gauss-Bonnett}
\newacronym{gc}{GC}{globular cluster}
\newacronym{gr}{GR}{general relativity}
\newacronym{grb}{GRB}{gamma ray burst}
\newacronym{gw}{GW}{gravitational wave}
\newacronym{gwb}{GWB}{gravitational wave background}
\newacronym{iad}{IAD}{integrated (ascribed) acceleration difference}
\newacronym{imr}{IMR}{inspiral-merger-ringdown}
\newacronym{isco}{ISCO}{innermost stable circular orbit}
\newacronym{js}{JS}{Jensen-Shannon}
\newacronym{kde}{KDE}{kernel density estimate}
\newacronym{kl}{KL}{Kullback-Liebler}
\newacronym{lfn}{LFN}{laser frequency noise}
\newacronym{ligo}{LIGO}{Laser Interferometer Gravitational-Wave Observatory}
\newacronym{lisa}{LISA}{Laser Interferometer Space Antenna}
\newacronym{mcmc}{MCMC}{Markov chain Monte Carlo}
\newacronym{mcs}{MCS}{maximum correlated samples}
\newacronym{mle}{MLE}{maximum likelihood estimate}
\newacronym{mpa}{MPA}{marginal precision and accuracy}
\newacronym{nsc}{NSC}{nuclear star cluster}
\newacronym{par}{PAR}{precision to accuracy ratio}
\newacronym{pcm}{PCM}{parameter covariance matrix}
\newacronym{pe}{PE}{parameter estimation}
\newacronym{pn}{PN}{post-Newtonian}
\newacronym{psd}{PSD}{power spectral density}
\newacronym{roq}{ROQ}{reduced order quadrature}
\newacronym{rpm}{RPM}{reduced parameter model}
\newacronym{smbh}{SMBH}{supermassive black hole}
\newacronym{snr}{SNR}{signal to noise ratio}
\newacronym{spa}{SPA}{stationary phase approximation}
\newacronym{tdi}{TDI}{time delay interferometry}
\newacronym{tt}{TT}{transverse traceless}
\newacronym{wss}{WSS}{wide sense stationary}

\title{Fast LISA Likelihood Approximations by Downsampling}

\author[1]{Jethro Linley \thanks{jethro.linley@glasgow.ac.uk}}
\affil[1]{School of Mathematics and Statistics, University of Glasgow, G12 8TA}
\begin{document}
\maketitle

\begin{abstract}

The Laser Interferometer Space Antenna (LISA) is due to launch in the mid-2030s. A key challenge for LISA data analysis is efficient Bayesian inference with parametrised gravitational-wave models, particularly for early inspirals of low- and intermediate-mass black-hole binaries, where time series can contain $\sim 10^8$--$10^9$ samples and naive likelihood evaluations become prohibitively expensive. For purely simulated studies, we present a simple time-domain likelihood-approximation scheme for such signals. The method retains only a small subset of samples and defines a modified noise-weighted inner product on this reduced set that closely reproduces the original inner product on the waveform manifold. Because this alters the effective noise model, the scheme is intended for simulated LISA data rather than direct analysis of real LISA data. In our examples, the resulting posteriors agree to high accuracy with those obtained using much denser sampling, while retaining only $10^3$--$10^4$ samples. The computational cost then scales linearly with the number of retained samples $\Ns$, so the speed-up is roughly $N/\Ns$, where $N$ is the original data length; for realistic LISA-like datasets with $N\sim10^8$--$10^9$ this would correspond to gains of order $10^4$--$10^6$ over straightforward frequency-domain likelihood evaluations of the full dataset. The time-domain formulation is particularly convenient for incorporating waveform modifications from non-trivial astrophysical environments or alternative-gravity effects. The results presented here provide the theoretical basis for the software package \texttt{Dolfen}.

\end{abstract}

\vspace{10pt}

\begin{multicols}{2}
\tableofcontents

\glsaddall

\section{Introduction}
Motivated by the need to perform Bayesian \gls{pe} on very large, purely simulated datasets with limited computing resources, we investigate a means of reducing the computational costs without jeopardising the accuracy of the results. Our approach will be to discard the majority of the data points and redefine a noise-weighted inner product on the retained data points (typically retaining only $\Ns \sim 10^3$--$10^4$ samples) to compensate for the discarded information. We shall refer to this procedure in this paper as \emph{downsampling}, and aim to quantify the robustness, limitations and benefits of downsampling using a given dataset, in particular such that the likelihood function used in Bayesian \gls{pe} is minimally affected. From this perspective, the task is an optimisation problem. Using fewer data points generally means fewer numerical operations are required to evaluate the likelihood function at a given point in parameter space. We will see that downsampling allows a drastic reduction in the likelihood computation time. 
The particular focus here is the study of `slowly evolving' \gls{gw} signals that we expect to obtain from the \gls{lisa} detector \cite{LISAwhitepaper} (with special emphasis on \emph{compact binary inspirals}), but the methods employed here are general and may be used with other datasets of signals that can be said to be slowly evolving in time.

The inspiral stage of \glspl{bhb} may last millions of years before merger, where the vast majority of the duration of \gls{gw} emission constitutes a very slowly evolving signal of gradually increasing amplitude and frequency (a characteristic `chirp' \cite{chirp}). The \gls{lisa} detector will be sensitive to a good portion of the lower frequency ($\sim$1--100\,mHz) part, and may be active for $\sim$5--10 years \cite{LISAwhitepaper}. For the most slowly evolving signals hovering near the upper frequency sensitivity of $\sim$100\,mHz throughout \gls{lisa}'s lifetime, we may realistically expect datasets consisting of $\sim 10^8$--$10^9$ samples at the Nyquist rate.

Such large datasets require substantial computing resources for \gls{pe}, but in preparatory work with fully simulated datasets we are free to use any approach that reproduces the desired information, including redefining aspects of the signal and/or detector model. Our sole objective is to obtain an accurate representation of the likelihood function as cheaply as possible, which downsampling achieves very effectively. This scheme is not intended for direct use with real LISA data, because it alters the effective noise model in a way that is only justified for simulated data with known noise properties; its use is therefore limited to fully simulated \gls{pe} environments. We use downsampling here as a fast yet high accuracy exploratory tool: to assess the viability of proposed analysis strategies, to quantify the waveform-modelling accuracy required for LISA, and to study the features of large-dataset likelihoods.

It is instructive to place this method and its results in the context of related PE acceleration approaches. Perhaps the most frequently employed formalism is the Fisher information approximation; see for example \cite{GrimmHarms, MBinGWTC1, FundPhyswGWs, ExpndLISAfrmGrnd, LISAEMRIsFisher, TransOrbRes}. This formalism provides a multivariate Gaussian approximation of the likelihood, giving a lower bound on the standard error of the parameters under the proper posterior. In many cases it is a poor approximation of the posterior; for a detailed discussion of why this is not generally reliable (except in some particular regimes), see \cite{UseAbuseFisher, InadequateFIM}. In follow-up work, we will show that this problem becomes severe for LISA likelihood functions of extended GW waveform models. 

Other methods that exploit common properties of the waveforms under study, by finding ways to cheaply produce waveform templates, include \emph{\gls{roq}} \cite{ROQ}, heterodyning/\emph{relative binning} \cite{cornish2013fastfishermatriceslazy, RelativeBinning}, and so-called \emph{\gls{afr}} \cite{AdaptFreqRes}. The related technique of \emph{Template-Interpolation} \cite{TemplateInterpolation} entails computing the frequency-domain waveform excluding those parts which are known to be roughly linear, and interpolating over the gaps between explicitly computed points. While \gls{afr} and template interpolation operate in the frequency domain, \gls{roq} relies on detector- and prior-specific offline training to construct reduced bases, and heterodyning makes very efficient use of a high-overlap reference waveform to compress the likelihood evaluation. Each of these approaches has regimes where it is highly effective, but they typically require problem-specific preparation (basis construction, training, or reference selection) when the waveform family, priors, or data model are changed.

Even in the presence of highly optimised heterodyned pipelines, a time-domain downsampling scheme remains useful. Heterodyning and relative binning achieve large gains by expanding around a high-overlap reference waveform, which makes them extremely efficient once a suitable reference has been chosen but also ties the compressed data product to that specific waveform family and signal model. By contrast, the downsampling developed here is defined entirely at the level of the data and the likelihood inner product: one selects a subset of samples with associated weights and then evaluates arbitrary time-domain waveforms only on that subset. The same reduced data and inner product can be reused for different waveform models and alternative-gravity signals, without regenerating model-dependent heterodyned streams each time the signal model changes. The accuracy of the approximation is controlled by a single parameter $\Ns$, which can be increased until the resulting posterior stabilises, as quantified in Section~\ref{sec:distcloseness}. In this sense, downsampling does not compete with heterodyning as a niche optimisation, but provides a generic, reusable data-compression layer that is valuable even when heterodyned likelihoods are available; there may even be scope to combine the two approaches by applying downsampling to heterodyned data products, though we shall not pursue this here.

A useful corollary is that nothing in the construction assumes a fixed number of sources. Once a sample subset and weight vector have been chosen for a given data stream, any composite waveform built from that stream is treated in exactly the same way: whatever configuration the sampler proposes is evaluated, whitened, and inserted into the same inner product on the retained samples. The computational bookkeeping associated with adding or removing sources is therefore decoupled from the data-compression step. We do not explore such applications in this work, but they provide a natural direction for future studies.

A demanding and increasingly routine task in GW astronomy is to quantify bias from waveform incompleteness~\cite{capuano2025, Kapil2024}. Signals are injected that include extra physics (e.g.\ eccentric periastron bursts, self-force terms, non-GR dispersion) and recovery is attempted with a baseline model that omits those effects. Each injection requires a full Bayesian run, and many are needed to chart bias versus SNR and source parameters. In such studies it is advantageous to use accelerators that do not require retraining or recalibration when the injected physics or recovery model are changed. This makes downsampling a natural candidate tool for large-scale bias surveys. The computational cost scales with the number of retained samples, and, where desired, the accuracy of the approximation can be checked by repeating the analysis for two or more values of $\Ns$ and confirming that the resulting posteriors are consistent within sampling noise. The implementation is minimal, compatible with gapped data, and integrates cleanly with standard samplers.

It is also appropriate to comment briefly on the choice of deriving this approximation for time-, rather than frequency-domain signals. A great deal of signal analysis, particularly in the field of \gls{gw} data analysis, is performed in the frequency domain; for stationary Gaussian noise the covariance is diagonal there, and FFT-based likelihood evaluations are correspondingly cheap. For a significant subset of the signals expected to be detected by LISA, however, parameterised models (or modifications thereof, from, for example, time-delays in the GW induced by galactic orbital motion of binaries and environmental effects) are straightforward to write down in the time domain, and conversely very difficult to formulate in the frequency domain. With this in mind, and the fact that some real LISA analyses are likely to take place in the time domain, it is very useful to have a time-domain method. This will also be useful for developing understanding of the intricacies of LISA time-series signal modelling and data analysis. It has recently been pointed out to the author that downsampling has been used in the recent work of Ref.~\cite{speri}. In that paper, the authors applied downsampling on strongly Gaussian posteriors of high-SNR EMRI waveforms in the frequency domain. In this paper, downsampling is somewhat more generalised, applying also to low-SNR signals with highly structured likelihood functions, and the time-domain formulation can often be very useful for constructing intricate and comprehensive waveform models.

In this work we formulate a generic time-domain downsampling scheme based on a modified noise-weighted inner product on a retained subset of samples; demonstrate on LISA-like inspiral signals that full posteriors can be recovered to high accuracy using only $\Ns \sim 10^3$--$10^4$ samples, with speed-ups of order $10^4$--$10^6$ over straightforward frequency-domain likelihood evaluations of the full dataset; introduce posterior-distance diagnostics to quantify the discrepancy between downsampled posteriors and to define practical convergence criteria; and provide an implementation in the software package \texttt{Dolfen} that fits directly into standard data analysis pipelines, enabling these techniques to be used in broader simulation studies.

This paper is structured as follows. In Section~\ref{sec:prelims} we introduce the framework used to define the problem and collect several results that will be required later. In Section~\ref{sec:newlikelihood} we define and develop different approaches to downsampling, discussing some of the concepts and issues that must be addressed to obtain a faithful reproduction of the \gls{lisa} posterior. Section~\ref{sec:testdownsamp} introduces numerical diagnostics for comparing high-dimensional posteriors and derives practical acceptance criteria for downsampling schemes; in that section we also describe test cases and compute and compare posteriors to evaluate the procedures experimentally. Section~\ref{sec:evaltimes} presents the projected improvements in the rate of evaluation of \gls{pe} attained by downsampling and summarises our main conclusions about the performance and limitations of the method. Finally, Section~\ref{sec:dolfen} describes the \texttt{Dolfen} software package, which implements the techniques developed here and is intended for use in broader simulation studies.

\section{Preliminaries}\label{sec:prelims}

\subsection{Slowly evolving waveforms}
Downsampling cannot be expected to work for an arbitrary parameterised waveform. An important requirement for a waveform to be amenable to downsampling is that it must be `slowly evolving'. We loosely define a waveform as \emph{slowly evolving} over a time interval $\mathcal{S}$ if, for each element of the Fisher information matrix (see Section~\ref{sec:approxmethods}), the average value over a moving window $\mathcal{I}_t$ centred at time $t$ is approximately constant for all $t \in \mathcal{S}$. Here $\mathcal{I}_t$ is an `intermediate-duration' interval whose length is of order $10^2$ times the sampling interval (i.e.\ the reciprocal Nyquist rate) of the waveform on $\mathcal{S}$. For the GW waveforms considered here, an equivalent characterisation is that the Fisher matrix elements, marginalised over coalescence phase, vary only slowly as a function of time sample across $\mathcal{S}$. A definition of this sort is needed to classify sinusoidal waveforms as slowly evolving when their frequency and amplitude parameters are roughly constant: although the waveform oscillates and the pointwise Fisher matrix elements oscillate with it, averaging over intermediate windows (or marginalising over phase) smooths out these oscillations.

To see why slow evolution is an important requirement for downsampling, consider the simplest time-series signal that is clearly slowly evolving: a constant signal. Taking data points from any location would provide equal information on the constant. Conversely, consider a signal with an instantaneous step from one constant to another at a certain time; one should clearly classify this as not slowly evolving. The data points just before and after the step provide the most information on the location of the step. Downsampling in this case will fail; removing any data point eliminates the ability to determine the original likelihood of the step location between the remaining samples, at the same resolution. Since we aim to introduce a general, signal-independent downsampling procedure, we restrict applications to signals of slow time evolution.

\subsection{Bayesian inference}\label{bayesinf}

If the data contain a signal described by a parametrised waveform model, the associated parameters can be inferred using Bayesian methods, as is now standard in \gls{gw} data analysis. The \emph{posterior} probability density for the parameters $\pvec$ given the data $\mathbf{d}$ is defined by Bayes' theorem,
\begin{equation}\label{bayes_thm}
p(\pvec\,|\,\mathbf{d}) = \frac{p(\pvec)\,p(\mathbf{d}\,|\,\pvec)}{p(\mathbf{d})} \, ,
\end{equation}
where $p(\pvec)$ is called the \emph{prior} probability of the parameters, written as the vector $\pvec$, $p(\textbf{d}\,|\,\pvec)$ is the \emph{likelihood}; the probability of obtaining data $\textbf{d}$ given parameters $\pvec$ and given a waveform model, $p(\mathbf{d})$ is the \emph{evidence} for the model given data $\textbf{d}$, and where the data is taken to be the sum, $\textbf{d}=\textbf{s}+\textbf{n}$, of the true signal, $\textbf{s}=\textbf{s}(\hpvec)$, and detector noise, $\textbf{n}$, where $\hpvec$ is the vector of signal parameters. We ignore the evidence here as we shall not be comparing different models.

For simplicity, we take uniform priors with a suitably chosen window centred on the true parameters, and appropriate sinusoidal priors for angular parameters. The remaining factor to be determined is the likelihood function. Assuming a Gaussian noise model, we can write the probability of some noise realisation $\mathbf{n}$ occurring (i.e., the likelihood) as
\begin{equation}\label{firstlikelihood}
p(\mathbf{n}) = \left[ (2\pi)^N\det(\cvm)\right]^{-1/2} \exp\left[-\tfrac{1}{2}\langle\mathbf{n}\,|\,\mathbf{n}\rangle\right] \, ,
\end{equation}
where $\cvm$ is the \emph{(noise) covariance matrix}, $N\equiv \dim(\mathbf{n})$, and ($\langle\cdot|\cdot\rangle$) is an inner product. In the time-domain this can be written \cite{Cornish_2020}
\begin{equation}\label{orig_td_inner_prod}
\langle \mathbf{a} | \mathbf{b} \rangle = {\mathbf{a}}^\T\cvm^{-1}{\mathbf{b}} \, .
\end{equation}

\section{Defining a new likelihood function}\label{sec:newlikelihood}

\subsection{Downsampling}

The first stage of downsampling is simply the process by which a dataset, $\ds$, say, consisting of $N_\s$ samples is defined, by some means or another, as a subset of an original dataset $\df$ which consists of $\Nf>\Ns$ samples; i.e., $\ds\subset\df$. For example, \emph{decimation} is a well-known and common method whereby every tenth sample is selected from some original ordered dataset, to form a new ordered dataset one tenth of the original's size.

\subsubsection{Sample selection schemes}\label{dsschemes}

After some preliminary investigation, there appears to be rather limited advantages to using specific downsampling \emph{schemes} (i.e., procedures for choosing a particular selection of data points for the subset). Some examples of schemes are:
\begin{itemize}
\item Uniform downsampling/`decimation' - taking every $n^\mathrm{th}$ sample ($n>1$).
\item Random downsampling - selecting $\Ns$ samples at random from the full dataset.
\item Cluster random sampling - random samples chosen from restricted regions of the dataset.
\item Hybrid sampling - combining uniform and random sampling; roughly half the samples taken uniformly, and half taken randomly from original dataset.
\end{itemize}
One sample selection scheme may allow recovery of a fully sampled posterior in fewer samples than some other scheme. Attempts to optimise the downsampling procedure using specific schemes should likely need to be studied for each type of model (even for each dataset considered); the efforts required for this will almost certainly outweigh any benefits and shall not be considered here. However, since we must choose some scheme, we will perform a comparison of various schemes using some fiducial system to decide, empirically, on the best performing scheme.

The scheme of greatest interest is random sampling, which is expected to give a good representation of the signal for high enough sampling density. Uniform (/periodic/regular) downsampling is expected to lead to aliasing and related effects, but by randomly sampling, this problem can be mostly eliminated \cite{Bretthorst}, so long as the number of samples used is not catastrophically low, causing the structure of the posterior to break down. In some preliminary tests, however, it was noticed that uniform downsampling is very effective: the structure of the posterior remains stable down to a surprisingly low sampling rate. We surmise that this might be due to the fact that, as a result of the non-white noise profile, the strain value of any given sample is informed by a number of its neighbouring samples, which are located at a distance so as to at least capture the Nyquist frequency components of the signal. Thus despite the vast downsampling of uncorrelated samples, some high frequency information remains since the decorrelation process encodes it in the residual.

Because uniform sampling performs well, we will examine the hybrid scheme, however, uniform sampling alone shall not be considered due to its potential to compromise the integrity of the likelihood in a general setting, by the reasons outlined above. The idea is that incorporating some degree of uniformity in the data point selection will ensure evenness in the relative spread of information across the signal and be less prone to random data point selection problems, securely allowing fewer samples to retain the required information, whilst effects such as aliasing are suppressed by the random samples. For these reasons we shall also examine the performance of cluster sampling.

\subsubsection{Residuals and metrics}\label{ResidualsAndMetrics}
There are a number of ways to express the downsampled residual in terms of the fully sampled residual. First, consider the $\Ns \times \Nf$ matrix
\begin{equation}\label{dsmat}
\mathbf{D}=\begin{pmatrix}
0 & 1 & 0 & 0 & 0 & 0 & 0 & 0 & \dots & 0 \\
0 & 0 & 1 & 0 & 0 & 0 & 0 & 0 & \dots & 0 \\
0 & 0 & 0 & 0 & 0 & 1 & 0 & 0 & \dots & 0 \\
0 & 0 & 0 & 0 & 0 & 0 & 1 & 0 & \dots & 0 \\
0 & 0 & 0 & 0 & 0 & 0 & 0 & 0 & \dots & 0 \\
\vdots & \vdots & \vdots & \vdots & \vdots & \vdots & \vdots & \vdots & \ddots & \vdots \\
0 & 0 & 0 & 0 & 0 & 0 & 0 & 0 & \dots & 0
\end{pmatrix} \, ,
\end{equation}
which, acting on the left of the $\Nf$ dimensional vector $\mathbf{v}=(v_1,v_2,v_3,v_4,v_5,\dots,v_{\Nf})^\T$ produces the $\Ns$ dimensional vector $\mathbf{v'}=\mathbf{Dv}=(v_2,v_3,v_6,v_7,\dots)^\T$. We shall endeavour to set the downsampled likelihood equal to the original likelihood, at least approximately. Equivalently, one can set the inner products approximately equal:
\begin{equation}
\mathbf{v'}^\T\xicvm_\s\mathbf{v}'\approx \mathbf{v}^\T\xicvm_\f\mathbf{v} \, ,
\end{equation}
for all $\mathbf{v}$, where $\xicvm_\s$ and $\xicvm_\f$ are inner products on the subspace and full data space respectively. In general, an exact solution for $\xicvm_\s$ would depend on the full residual vector $\mathbf{v}$. Our method will be to make an initial guess of a metric on the subspace ${\xcvm'}_\s^{-1}$, and define the final metric $\xicvm_\s=\xicvm_\s(\xcvm'_\s)$ in terms of the initial guess in some appropriate fashion, as shall be described in the following sections. With ${\xcvm'}_\s^{-1}$ an $\Ns\times\Ns$ matrix, we can project the full data space metric down to the subspace using $\mathbf{D}$ via
\begin{equation}
{\xcvm'}_\s^{-1} = \mathbf{D}\xicvm_\f\mathbf{D}^\T \, .
\end{equation}
As an example, we will later try to find some constant $m$ to set $\xicvm_\s=m{\xcvm'}_\s^{-1}$ (see Section \ref{sec:gen_gauss_approx}) in which case one may write
\begin{equation}\label{approxexample}
\mathbf{v}^\T\xicvm_\f\mathbf{v} \approx \mathbf{v'}^\T\xicvm_\s\mathbf{v}'=m\mathbf{v}^\T\mathbf{D}^\T\mathbf{D}\xicvm_\f\mathbf{D}^\T\mathbf{D}\mathbf{v} \, .
\end{equation}
Note another equivalent way of expressing the downsampling operation: using a square matrix $\mathbf{S}$ to select samples from $\mathbf{v}$, such that $\mathbf{v}'$ remains $\Nf$-dimensional, i.e, the identity matrix but with some diagonals set to zero, e.g.: $\mathbf{S}=\mathrm{diag}(0,0,1,0,...,0)$. This is equivalent to the above since such a matrix may be defined as $\mathbf{S}=\mathbf{D}^\T\mathbf{D}$.

A further useful way to downsample, rather than reducing the number of dimensions by projection using $\mathbf{D}$, is to define an $\Nf$-dimensional \emph{sample selection vector}, $\mathbf{k}=(0,1,1,0,0,1,1,0,\dots,0)$, and write $\mathbf{v}'=\mathbf{v}\odot\mathbf{k}$, where $\odot$ is the Hadamard (element-wise) product, then simply using $\xcvm'_\s=\xcvm_\f$ as the initial guess. This retains the dimensionality and `deletes' the data we do not intend to use; the usefulness of this approach will become apparent in following sections.

\subsubsection{Time-domain sample correlations}\label{timedomaincorr}

Suppose that the acceptable optimal downsampling rate is very high: the remaining few samples will, on average, be greatly separated in time. In this case, correlations between the remaining samples would be effectively vanishing and can be ignored. One might therefore suppose the remaining samples can be treated as having effectively been produced by a detector with a white noise profile. Whilst it is true that these samples are effectively not correlated with each other, they will, in the full \gls{lisa} dataset, be highly correlated with their immediately neighbouring samples. In the full \gls{lisa} dataset analysis, these correlated neighbours are of course not neglected, and inform the likelihood function.

It is not obvious that neglecting correlated neighbouring samples will cause a significant deviation from the true posterior, but our preliminary posterior convergence validity tests showed that it is generally unsafe to do so. Upon reflection, the effect is quite reasonable and may be explained with a simple example. Consider estimating the phase of a dirac-comb signal of known amplitude and frequency from a detector with a white noise profile (where the autocorrelation function is a delta function). Now, compare this to estimating the phase of the same signal measured by another detector with a noise profile such that immediately neighbouring samples are highly anti-correlated; in this case, one has more freedom to shift the phase from the truth by one sample without changing the Mahalanobis distance (and so the posterior) as much, since the residual in that case is more likely to be attributable to noise only. Hence in the downsampled case, if one selects a sample without considering how its value is affected by its neighbours, one will likely arrive at the wrong conclusions about how it constrains the model parameters.

The above example shows how correlations cause the information in the time-domain signal to which the detector is sensitive to become mixed between neighbouring samples. In order to prevent this from impinging upon the Bayesian inference after downsampling, we must whiten, or in some other way decorrelate the data samples (for example, frequency representations from \gls{fft}s often consist of uncorrelated samples). Thus in fact the preceding prescription for downsampling given in Section \ref{ResidualsAndMetrics} \emph{should only be carried out on uncorrelated data}: since frequency-domain data is naturally uncorrelated, the task would be straightforward and safe (at least from the point of view of sample correlations) to downsample frequency samples. In the time-domain, we must first whiten residuals before downsampling. This simply means setting
\begin{equation}\label{whtdsig}
\mathbf{v}=\cvm^{(-1/2)}\mathbf{r} \, ,
\end{equation}
where $\mathbf{r}$ is the time-domain residual and $\mathbf{v}$ its whitened counterpart, so that the inner product of time-domain residuals can be written
\begin{equation}
\langle \mathbf{r \,|\, r} \rangle = \mathbf{r}^\T\icvm\mathbf{r} = (\cvm^{(-1/2)}\mathbf{r})^\T\cvm^{(-1/2)}\mathbf{r} = \mathbf{v}^\T\xicvm\mathbf{v} \, ,
\end{equation}
where the metric in the `whitened basis' $\xicvm=\mathds{1}$.

Whitening time-domain signals via the matrix operation in (\ref{whtdsig}) is generally extremely costly. However, due to being able to downsample our data to a high degree, an efficient algorithm is possible for computing only the required template samples (to within some chosen degree of accuracy regarding the number of `significant' neighbouring samples, see Section \ref{sec:MCS} below) to produce the downsampled residual
\begin{equation}\label{dsresidual}
\mathbf{v}\odot\mathbf{k}=\cvm^{(-1/2)}\mathbf{r}\odot\mathbf{k} \, ,
\end{equation}
where $\mathbf{k}$ is a sample selection vector. If it can be shown that the approximation method where we find the optimal noise reduction factor $m$ given in (\ref{approxexample}) is accurate, for example, then we can write the approximate inner product as
\begin{align}\label{newinnerprod}
\langle \mathbf{r \,|\, r} \rangle &\approx m(\mathbf{v}\odot\mathbf{k})^\T(\mathbf{v}\odot\mathbf{k})\, .
\end{align}

\subsubsection{Maximum correlated samples}\label{sec:MCS}

The downsampled residual defined in (\ref{dsresidual}) requires an expensive matrix multiplication to evaluate. As we will see later, we are generally allowed a definition of $\mathbf{k}$ that vastly reduces the number of signal samples that are required to be calculated to compute $\mathbf{v}\odot\mathbf{k}$. However, for each `1' in $\mathbf{k}$ corresponding to a selected \emph{decorrelated} sample, we are still required to know a few more samples from $\mathbf{r}$ (the `significantly correlated' samples) to be able to perform the decorrelation and compute the element of $\cvm^{(-1/2)}\mathbf{r}$ corresponding to the relevant `1' in $\mathbf{k}$.

The number of elements of $\mathbf{r}$ required to be computed for each `1' in $\mathbf{k}$ depends on the \gls{lisa} \gls{acf}. The \gls{acf} always tends to zero for large values of the time lag. In practice however, an \gls{acf} tends to zero even for relatively small values of the time lag, so if we set some level of precision beyond which correlations can be ignored, the number of elements of $\mathbf{r}$ required to compute an element of $\mathbf{v}\odot\mathbf{k}$ can be reduced significantly. For our analysis, we used the level of accuracy given by:

\begin{definition*}
The (approximate number of) \emph{\gls{mcs}}, denoted $\mathcal{M}$, is the smallest index of the cumulative sum of the absolute value of the (one-sided) whitening function (CSAWF) for which the cumulative sum exceeds 97\% of its asymptotic value.
\end{definition*}

For any given signal--detector model, this definition ensures that at least 97\% of the contribution to a given whitened residual sample is explicitly accounted for. Some cutoff of this kind is essential: without it, one would in principle need to compute all data points to perform the pre-whitening, which would defeat the purpose of downsampling. For realistic detector noise profiles, whose autocorrelation functions decay smoothly, the remaining few percent of the whitening weight is concentrated at large lags and is therefore expected to contribute negligibly to the residual value.

Any given signal has its own Nyquist rate. This fixes the sampling interval (and hence the frequency-bin size), which in turn determines the resolution of the \gls{acf}; the \gls{mcs} is therefore generally different for each signal. If some system has an \gls{mcs} of $\mathcal{M}=2$, then $2\mathcal{M}+1 = 5$ elements of $\mathbf{r}$ must be known for each `1' in $\mathbf{k}$: the sample corresponding to the `1' and its two immediate predecessors and successors. For example, if $k^{76}=1$, then we require $r^{74},\dots,r^{78}$ in order to compute $v^{76}$ accurately. In general, $2\mathcal{M}+1$ elements of $\mathbf{r}$ must be computed for each `1' in $\mathbf{k}$.

This counting is only approximate, however, because the required neighbourhoods can overlap. For instance, if also $k^{77}=1$, then we nominally need $r^{75},\dots,r^{79}$, but $r^{75},\dots,r^{78}$ have already been obtained from the requirements of $k^{76}=1$. We denote the approximate total number of template samples that must be evaluated by $N_\mathcal{M}$, which for $\Ns \ll \Nf$ is well approximated by
\begin{equation}\label{MCSsamples}
    N_\mathcal{M}(\Ns) \approx (2\mathcal{M}+1)\Ns \, .
\end{equation}

\subsection{Likelihood approximation methods}\label{sec:approxmethods}
To reproduce the effect of the full-data likelihood under downsampling, we require a way to approximate the likelihood evaluated on the reduced dataset. Our approach is to start from a natural downsampled likelihood and then `reshape' it towards the full-data likelihood by correcting for the change induced by discarding samples. We attempt this correction by matching, at least in part, the second-order structure of the posterior. In the Gaussian/high-SNR regime, this is encoded by the parameter covariance matrix (or equivalently the Fisher information matrix). In one approach we adjust a single noise-reweighting factor so that a scalar summary of this structure (the overall volume of the covariance ellipsoid) is preserved under downsampling. In a more general approach we take the full Fisher matrices into account.

For intuition, let us first consider the posterior defined by data consisting of a signal plus a zero-noise realisation. In this case the residual $\mathbf{r}(\pvec)$ vanishes at the true parameters, so the maximum-likelihood point is unchanged by downsampling: modifying the inner product or discarding samples does not move the point where $\mathbf{r} = 0$. The leading effect of downsampling is therefore on the curvature of the log likelihood around this maximum, i.e.\ on the second-order structure of the posterior.

In the Gaussian/high-SNR regime this second-order structure is encoded in the parameter covariance matrix (PCM), whose entries are the second central moments of the posterior. The PCM is well approximated by the inverse of the \emph{\gls{fim}}, given by \cite{FisherDef, FisherDef2}:
\begin{equation}\label{firstFIM}
F_{ij}\equiv \left[\frac{\partial h^\T}{\partial \theta_i}\icvm\frac{\partial h}{\partial \theta_j}+\frac{1}{2}\mathrm{tr}\left(\icvm\frac{\partial\cvm}{\partial\theta_i}\icvm\frac{\partial\cvm}{\partial \theta_j}\right)\right]_{{\pvec}=\hpvec} \, ,
\end{equation}
which is evaluated here at $\hpvec$, the maximum likelihood estimate. For stationary noise (where $\cvm$ is constant) this reduces to
\begin{equation}\label{FIM}
F_{ij}\equiv \left[\frac{\partial h^\T}{\partial \theta_i}\icvm\frac{\partial h}{\partial \theta_j}\right]_{\pvec=\hpvec} \, .
\end{equation}
It is useful to abbreviate this, dropping the evaluation point instruction and taking this to be implied, writing
\begin{equation}\label{FIMdef}
F_{ij}=h_{k,i}\icvm_{kl}h_{l,j} \, ,
\end{equation}
where summing over repeated indices is assumed, and where $h_{k,i}\equiv\frac{\partial}{\partial\theta^i}h_k$. 

\subsection{A single noise reduction factor}

We conjecture that a single noise reweighting factor applied to all remaining samples may be sufficient to recover the general structure of the distribution after downsampling, provided enough samples remain to capture a fair representation of the information content.

To motivate this conjecture, consider the simple model $\mathbf{h}(\theta) = \theta \mathbf{1}_\f$, where $\mathbf{1}_\f$ is an $\Nf$-dimensional vector of ones. Suppose the detector produces time series $\df = \Sf + \mathbf{n}_\f$, with $\Sf = \mathbf{h}(\hat{\theta}) = \hat{\theta}\,\mathbf{1}_\f$, where $\hat{\theta}$ is the true parameter, and $\mathbf{n}_\f$ is a realisation of white Gaussian noise with covariance $\mathbb{C}_\f = \sigma_\f^2 \mathds{1}$. The residual is
\[
\mathbf{r}_\f(\theta) = \df - \mathbf{h}_\f(\theta) = \mathbf{n}_\f + (\hat{\theta}-\theta)\mathbf{1}_\f \, .
\]
Ignoring additive constants, the log likelihood is
\[
\ell_\f(\theta) \propto -\frac{1}{2\sigma_\f^2}\,\mathbf{r}_\f^\T(\theta)\,\mathbf{r}_\f(\theta) \, .
\]
Expanding the norm gives
\[
\mathbf{r}_\f^\T \mathbf{r}_\f
= \mathbf{n}_\f^\T\mathbf{n}_\f
  + (\hat{\theta}-\theta)^2 \mathbf{1}_\f^\T\mathbf{1}_\f
  + 2(\hat{\theta}-\theta)\,\mathbf{1}_\f^\T\mathbf{n}_\f \, .
\]
For zero-noise data or upon averaging over noise realisations, the cross term vanishes and $\mathbf{1}_\f^\T\mathbf{1}_\f = \Nf$, so we obtain
\[
\ell_\f(\theta) \approx -\frac{\Nf}{2\sigma_\f^2}(\theta-\hat{\theta})^2 \, .
\]

Performing the same construction for a downsampled dataset of size $\Ns$ with white noise variance $\sigma_\s^2$, we find
\[
\ell_\s(\theta) \approx -\frac{\Ns}{2\sigma_\s^2}(\theta-\hat{\theta})^2 \, .
\]
To match the curvature of the two log likelihoods we can choose
\[
\sigma_\s = \sigma_\f \sqrt{\frac{\Ns}{\Nf}} \, ,
\]
so that $\ell_\s(\theta) \approx \ell_\f(\theta)$.

We can try to extend this sort of analysis to the signals and detector noise profiles of interest. However, the complexity increases very quickly for realistic physical systems, and closed-form modifications become unfeasible. It is also desirable to have a more general procedure that is model- and noise-profile independent.

Two (model-independent) noise reduction factors are derived in the following subsections, in which we approximate posteriors as Gaussian distributions, minimise the variation of certain scalar summaries of their second-order structure, and apply the corresponding modification to the downsampled posterior. In the high-\gls{snr} limit, where posteriors closely approximate Gaussians, this should be a good approximation (provided the semi-axes of the Gaussians are reasonably well aligned and proportionately similar). In other words, we expect that, for our restricted class of signals of interest (\gls{lisa} inspiral-only signals), downsampling mainly affects the overall \emph{precision} of the likelihood rather than its shape.

Since a \gls{pcm} describes the aspect ratio of the $k$-dimensional ellipsoid associated with the 1-sigma contours of the distributions \cite{UseAbuseFisher}, one might already guess that if the two ellipsoids are similar in orientation and aspect ratio, then the appropriate noise reduction factor is the one that rescales their volumes to agree. This is indeed the case; details are given in Section~\ref{primitiveFactor}. For a more robust treatment in Section~\ref{sec:gen_gauss_approx}, we also compute a factor that accounts for the Gaussians being misaligned and/or having unequal aspect ratios.

\subsubsection{PCM volume method}\label{primitiveFactor}

Here we assume that the semi-axes of the downsampled \gls{pcm} (and hence of the downsampled posterior) have the same orientations and aspect ratios as those of the full-data \gls{pcm}. The corresponding \gls{fim}s, $\fish'_\s$ and $\fish_\f$, then differ only by an overall scaling factor, so that
\[
m_\mathrm{det}\,\fish'_\s = \fish_\f \, .
\]
Taking determinants gives
\begin{equation}\label{mfactor}
m_\mathrm{det}=\sqrt[\leftroot{-1}\uproot{6}\scriptstyle k]{\frac{\det \fish_\f}{\det \fish'_\s}} \, ,
\end{equation}
where $k$ is the number of parameters. For some parameters, very little Fisher information exists, which can lead to numerical issues; these directions must be projected out of the parameter space before computing $m_\mathrm{det}$.

\subsubsection{General Gaussian approximations}\label{sec:gen_gauss_approx}

Without assuming that the downsampled \gls{fim} is aligned and proportionately similar to the full-data \gls{fim}, we instead derive a noise reduction factor for the subspace inner product by minimising the so-called \emph{Jeffreys' divergence} between posteriors approximated by concentric but otherwise general $k$-dimensional Gaussians; see Appendix~\ref{app:JeffDiv} for the derivation and details. The resulting noise reduction factor, from equation~(\ref{app:Jeff_div_min}), is
\begin{equation}\label{Jeff_div_min}
m_\mathrm{J}= \sqrt{\sum_{i=1}^k\frac{D^{-1}_{\s,ij}F_{\f,jl}D_{\s,li}}{D^{-1}_{\s,ij}F'_{\s,jl}D_{\s,li}} \cdot \sum_{i=1}^k\frac{D^{-1}_{\f,ij}F_{\f,jl}D_{\f,li}}{D^{-1}_{\f,ij}F'_{\s,jl}D_{\f,li}}} \; ,
\end{equation}
where $\mathbf{D}_\s$ diagonalises $\fish'_\s$ and $\mathbf{D}_\f$ diagonalises $\fish_\f$. Again, parameters with very little Fisher information must be projected out to prevent numerical issues. In preliminary tests we found that $m_\mathrm{J} \sim m_\mathrm{det}$ whenever the number of retained data points was not extremely small; we shall nevertheless use $m_\mathrm{J}$, since it arises from an optimisation based on a standard notion of distance between distributions.

To recap, we define the downsampled log likelihood using $m_\mathrm{J}$ as
\begin{equation}\label{noise_fac_llhood}
    \ell = -\tfrac{1}{2}m_\mathrm{J} (\bar{\mathbf{r}}\odot\mathbf{k})^\T(\bar{\mathbf{r}}\odot\mathbf{k}) \, ,
\end{equation}
where $\bar{\mathbf{r}}$ is the whitened residual and $\mathbf{k}$ is the sample selection vector.

\subsection{Preserving Fisher information}\label{preserveFIM}
There exist Fisher-information-lossless encodings of the data under downsampling, as we show in Appendix~\ref{app:pres_fish}; that is, the \gls{fim} can be preserved exactly after downsampling. In this construction the downsampled log likelihood is written as in equation~(\ref{fim_pres_llhood}), reproduced here:
\begin{equation*}
    \ell = -\tfrac{1}{2} (\mathbf{D}\bar{\mathbf{r}})^\T\xicvm_\s(\mathbf{D}\bar{\mathbf{r}}) \, ,
\end{equation*}
where $\bar{\mathbf{r}}$ is the whitened residual, $\mathbf{D}$ is the sample selection matrix, and $\xicvm_\s$ is the inner-product matrix on the whitened, downsampled signal space, as derived in Appendix~\ref{app:pres_fish}.

\subsection{Summary}

The downsampling procedure we have devised, though not necessarily implemented exactly as described here (for reasons of computational efficiency and algorithmic practicality), consists of the following steps:
\begin{enumerate}
\item Compute the \gls{fim} for the full \gls{lisa} posterior and the whitening matrix $\cvm^{(-1/2)}$ from the \gls{psd}.
\item Generate an $\Nf$-dimensional sample-selection vector $\mathbf{k}=(0,0,0,1,0,1,0,0,1,0,0,\dots)$ containing $\Ns$ ones and $(\Nf-\Ns)$ zeros, randomly positioned (or equivalently, construct a matrix $\mathbf{D}$ as in (\ref{dsmat})).
\item Use $\mathbf{k}$ to define the (whitened) downsampled residual,
      \[
      \bar{\mathbf{r}} = \cvm^{(-1/2)}\mathbf{r}\odot\mathbf{k} \, .
      \]
      This representation is convenient for efficient computation given the whitening step. An $\Ns$-dimensional version, $\bar{\mathbf{r}}_{\Ns}=\mathbf{D}\cvm^{(-1/2)}\mathbf{r}$, is conceptually equivalent and may be clearer for visualising the downsampling.
\item Define a new inner product on residuals and hence a new (log-)likelihood, as in either (\ref{noise_fac_llhood}) or (\ref{fim_pres_llhood}). This is constructed to be approximately equal to the (log-)likelihood defined with the full dataset for all $\pvec$, and its performance will be tested in the following section.
\end{enumerate}

\section{Posterior Production and Analysis}\label{sec:testdownsamp}

Intuitively, fewer samples mean cheaper likelihood evaluations and faster \gls{pe} convergence; this is the main motivation for downsampling. If $\Ns \approx \Nf$ (e.g.\ dropping only a small fraction of samples from a dataset $\df$ with $\dim(\df)\approx 10^7$), the downsampled posterior will be essentially indistinguishable from the full \gls{lisa} posterior (provided the retained-sample noise is adjusted appropriately), but the speed-up is modest. As $\Ns$ is reduced more aggressively, the remaining data become consistent with qualitatively different parameter combinations, producing extra modes and more highly structured posteriors. For example, uniform decimation below the Nyquist rate leads to aliasing. Random sampling, rather than regular decimation, largely mitigates this, but at very low $\Ns$ the posterior can still `break down' in the sense of differing qualitatively and unacceptably from the full-data posterior. Our goal is therefore to quantify the onset of this breakdown and to determine the smallest $\Ns$ that still reproduces the true posterior faithfully.

Any change to the data or likelihood definition (such as downsampling) inevitably alters the posterior, and it is not a priori clear when such changes should be regarded as `acceptable', since this depends on the application and on which features of the posterior are scientifically relevant. We therefore tailor our acceptance criteria to the LISA BHB \gls{pe} environment of interest here. Before introducing these criteria in Section~\ref{sec:convcrit}, we first describe sources of variation, or `posterior noise', in Section~\ref{sec:postnoisesources}, and then introduce a quantitative measure of posterior distance in Section~\ref{sec:distcloseness}.

\subsection{Posterior distribution noise}\label{sec:postnoisesources}

The following are brief qualitative descriptions of sources of variation present in posterior distribution estimates, the statistical measures will be presented in Section \ref{sec:convcrit}.

\subsubsection{Sampler noise}

It is generally infeasible to evaluate the posterior on a dense grid in parameter space for realistic dimensionalities, so we rely on stochastic samplers to construct approximate posterior estimates \cite{emcee, CPnest, Dynesty}. These estimates can vary slightly between different samplers, between different settings of a given sampler, and between different realisations of the underlying random processes. We compared the \gls{mcmc} sampler \texttt{emcee} \cite{emcee} and the nested samplers \texttt{CPnest} \cite{CPnest} and \texttt{Nessai} \cite{Nessai}, and found good agreement with our chosen default, \texttt{Nessai}. With fixed settings, repeated \texttt{Nessai} runs on the same data and likelihood differ only through the sampler's internal randomness. We refer to this irreducible variation as \emph{sampler noise}; it induces a noise floor that cannot be overcome.

\subsubsection{Downsampling noise}
Each data sample carries different information about the model parameters. If $\Ns$ samples are selected from an original set of $\Nf$ samples with $\Ns \gg 1$ (and the signal model is slowly evolving), the selected subset is likely to be a good representative of the information in the full dataset. Nonetheless, the geometry will vary slightly with the precise choice of the $\Ns$ samples. Thus different random data point selections (i.e.\ different realisations of $\mathbf{k}$ or $\mathbf{D}$) induce some variation in the posterior, which we refer to as \emph{downsampling noise}. As $\Ns$ is decreased, a typical selection becomes less likely to provide a good representation of the information present in the original dataset.

\subsubsection{Physical noise}

Real data contain detector noise. Different realisations of this noise can significantly alter the posterior; the dominant effect (as measured by divergence from the zero-noise posterior) is typically a shift in the position of the \gls{mle}, while changes in higher moments contribute smaller variations.

To quantify the intrinsic level of posterior-to-posterior variation induced by physical noise, we consider an ensemble of posteriors obtained from multiple independent noise realisations for the same injected signal and likelihood. For each realisation we \emph{mean-zero} the posterior in parameter space. This removes the overall shifts and focuses on changes in the posterior \emph{shape}. We then compute the distribution of distances between these mean-zeroed posteriors using the divergence measures introduced below. This defines a ``physical-noise floor'': the typical amount of variation in posterior geometry that arises purely from changing the detector noise realisation.

The resulting noise floor is system- and \gls{snr}-dependent, and is therefore tied to the specific LISA BHB configurations studied here. Further, the precise value depends on the chosen alignment prescription (here, mean-zeroing in each parameter); this choice is fixed throughout. Within these conventions, we treat downsampling-induced distortions as acceptable whenever the corresponding posterior distances are comparable to, or smaller than, this physical-noise floor.

\subsection{Approximating closeness of distributions}\label{sec:distcloseness}

The Kullback-Leibler (KL) divergence is a well-known (non-symmetric) measure
of discrepancy between two distributions, defined by
\begin{equation}\label{KLdivdef}
  D_{\mathrm{KL}}(p(\boldsymbol{\theta})\,||\,q(\boldsymbol{\theta}))
  \equiv
  \int_\Theta \mathrm{d}^k\theta \,
  p(\boldsymbol{\theta})
  \ln\!\left(\frac{p(\boldsymbol{\theta})}{q(\boldsymbol{\theta})}\right),
\end{equation}
where \(p\) and \(q\) are distributions of the continuous variable
\(\boldsymbol{\theta}\in\Theta\), with \(\Theta\) the parameter space.
Typically, \(p\) is viewed as the target/“true” distribution, and \(q\) as an
approximate or hypothesised distribution.

The Jensen-Shannon (JS) divergence is defined as
\begin{align*}\label{JSdivdef}
  D_{\mathrm{JS}}\Big(p(\boldsymbol{\theta})\,\Big|\Big|\,q(\boldsymbol{\theta})\Big)
  \equiv\;&
  \tfrac{1}{2} D_{\mathrm{KL}}\Big(p\,\Big|\Big|\,\tfrac{1}{2}[p+q]\Big)
  \\
  &+ \tfrac{1}{2} D_{\mathrm{KL}}\Big(q\,\Big|\Big|\,\tfrac{1}{2}[p+q]\Big)
  \numberthis
\end{align*}
and is a symmetric, bounded divergence between distributions. Moreover, its square root, \(\sqrt{D_{\mathrm{JS}}(p||q)}\), defines a metric on the space of distributions. This will be better suited for our requirements, particularly in situations where the exact target distribution is unattainable.

As mentioned, the posterior distributions themselves are computationally very expensive to evaluate accurately, and for high-dimensional parameter spaces the exact divergences are more expensive still: one must first estimate the posteriors, and then perform the additional integrals implied by the definitions above. We therefore resort to approximation techniques in order to obtain practical measures of the closeness between posteriors.

\subsubsection{Combined distances of marginal distributions}

Consider using divergences between \emph{marginal} distributions. Marginalisation is not ideal; being a projection, it necessarily discards information, but it remains useful as a low-cost diagnostic: if the 1D marginals of two posteriors become extremely similar, this strongly suggests convergence, and divergences between 1D distributions are relatively inexpensive to compute. There are infinitely many ways in which to project a distribution with $\geq 2$ dimensions into 1-dimension (by parameter mixing). We are usually interested in the unmixed model parameter constraints however, so it is natural to choose the projections that are the marginalisations onto the unmixed model parameters.

The Jensen-Shannon divergence \(D_\mathrm{JS}(p\Vert q)\) induces a metric
\[
  d_\mathrm{JS}(p,q) \equiv \sqrt{D_\mathrm{JS}(p\Vert q)}
\]
on the space of 1D distributions. We now combine these marginal distances into a single quantity by taking a weighted Euclidean norm:
\begin{equation}\label{marginalisemetric}
  \overline{d}_\mathrm{JS}(p,q)
  \equiv
  \sqrt{\sum_{i=0}^{k-1} 
    W_i \, d_\mathrm{JS}^2\!\left( O^i(p), O^i(q)\right)} ,
\end{equation}
where the operator
\begin{equation}
  O^i(p)
  \equiv
  \int_{\hat{\Theta}_i} p(\boldsymbol{\theta}) \,
  \mathrm{d}\theta^1 \cdots \mathrm{d}\theta^{i-1}
  \mathrm{d}\theta^{i+1}\cdots\mathrm{d}\theta^{k}
\end{equation}
projects the distribution \(p\) onto the \(i\)-th parameter axis, and
\(\hat{\Theta}_i \cong \mathbb{R}^{k-1}\) is the space spanned by
\(\hat{\boldsymbol{\theta}}_i=(\theta^1,\ldots,\theta^{i-1},\theta^{i+1},\ldots,\theta^{k})\).

We define the weights
\begin{equation}
  2 W_i \equiv
  \frac{\mathcal{H}_{O^i(p)}}{\sum_{j=0}^{k-1}\mathcal{H}_{O^j(p)}} +
  \frac{\mathcal{H}_{O^i(q)}}{\sum_{j=0}^{k-1}\mathcal{H}_{O^j(q)}} \, ,
\end{equation}
where \(\mathcal{H}_\mathcal{X}\) denotes the (Shannon) entropy of the
distribution \(\mathcal{X}\). Thus each marginal contributes to the combined
distance according to its relative contribution to the total marginal entropy
of \(p\) and \(q\). With this choice we have \(\sum_i W_i = 1\).

Since \(d_\mathrm{JS}\) is a metric on each marginal and \(\overline{d}_\mathrm{JS}\)
is a weighted \(\ell^2\) combination of these marginal distances, it defines a
metric on the space of \(k\)-tuples of 1D marginals. When regarded as a
function of the full joint posteriors \(p\) and \(q\), it should formally be
viewed as a pseudometric, because distinct joint distributions may share the
same set of marginals. This limitation is acceptable for our purposes, as we
use \(\overline{d}_\mathrm{JS}\) purely as a convergence diagnostic based on
parameter-wise constraints.

Squaring the combined distance yields a corresponding divergence
\begin{equation}\label{margdiv}
  \overline{D}_\mathrm{JS}(p\Vert q)
  \equiv
  \sum_{i=0}^{k-1} W_i \,
  D_\mathrm{JS}\!\left(O^i(p)\Vert O^i(q)\right) ,
\end{equation}
which we refer to as the \emph{combined marginal JS (CMJS)} divergence. With
logarithms taken in base~2, each \(D_\mathrm{JS}\) lies in \([0,1]\), and the normalisation \(\sum_i W_i = 1\) ensures that 
\(\overline{D}_\mathrm{JS} \in [0,1]\) as well. Analogously, we define the \emph{combined marginal Kullback-Leibler (CMKL)} divergence as
\begin{equation}
  \overline{D}_\mathrm{KL}(p\Vert q)
  \equiv
  \sum_{i=0}^{k-1} W_i \,
  D_\mathrm{KL}\!\left(O^i(p)\Vert O^i(q)\right) .
\end{equation}
These combined marginal divergences will be used in Sec.~\ref{sec:convcrit} as practical measures of the difference in posterior geometry under changes in downsampling rate.

\subsection{Acceptance criteria}\label{sec:convcrit}

We now propose criteria for determining the minimum number of samples required to define an acceptable downsampled posterior. There is no pre-existing benchmark, and different applications may tolerate different levels of approximation. Here we lay out a particular choice, suitable for recreating the LISA BHB \gls{pe} test environment up to an accuracy in posterior geometry set by physical noise, which is in line with our purpose for this work. Recall that we have three types of posterior noise to consider: physical noise, sampler noise, and downsampling noise. We first define how to quantify noise between distributions in Section~\ref{sec:ind_dist_noise}, use posterior variability under the sampler and downsampling to ``bootstrap'' an acceptable approximate target posterior in Section~\ref{sec:bootstrapping}, and then use physical noise to set an accuracy threshold in Section~\ref{sec:acc_thresh}.

\subsubsection{Induced distribution noise}\label{sec:ind_dist_noise}
Using the approximate divergences outlined in the previous section, we can begin to characterise basic statistical properties of the variation in the closeness of distributions arising from a given noisy process (i.e.\ the different sources of posterior noise). We now define two one-parameter (in $\Ns$) families of sets of divergences:

\begin{enumerate}
    \item When a target distribution is not known, the JS divergence is useful as a notion of relative distance. Given $K$ posterior estimates, we can form a population of $\binom{K}{2}$ pairs of posteriors and thus a set of $\binom{K}{2}$ divergences. We define
    \begin{equation}\label{setX}
        X_\mathrm{est}(\Ns)
        = \left\{
        \overline{D}_\mathrm{JS}\!\left(p_\mathrm{est}^{(i)}\,\Big|\Big|\,p_\mathrm{est}^{(j)}\right)
        : 1 \le i < j \le K
        \right\},
    \end{equation}
    where $p_\mathrm{est}^{(i)}$ is the $i^\mathrm{th}$ estimated posterior.
    \item When a target distribution is known, we have a notion of absolute distance in which the target plays the role of an “origin’’ in divergence space. With $K$ posterior estimates and a target distribution $q$, we define the set of $K$ divergences
    \begin{equation}\label{setY}
        Y^q_\mathrm{est}(\Ns)
        = \left\{
        \overline{D}_\mathrm{KL}\!\left(q\,\Big|\Big|\,p_\mathrm{est}^{(i)}\right)
        : 1 \le i \le K
        \right\}.
    \end{equation}
\end{enumerate}
Let the mean of a set of divergences $X$ be denoted $\mu[X]$, and the standard
deviation by $\sigma[X]$. The empirical distributions of these divergences are strictly non-negative and typically skewed, so a Gaussian approximation is not exact; nevertheless, $\mu[X]$ and $\sigma[X]$ provide convenient summary statistics with which to quantify “distribution noise’’ and to formulate consistent acceptance criteria for downsampled posteriors.

\subsubsection{Bootstrapping}\label{sec:bootstrapping}

Using fully sampled posteriors as target distributions for assessing the accuracy of downsampled posteriors is not feasible in realistic applications, because obtaining the former is prohibitively expensive. We therefore introduce a bootstrapping procedure to demonstrate that our approximations are sufficiently accurate and to define a minimal reliable downsampling rate.

The basic idea is to regard the variability of posteriors under downsampling as a function of \(N_\s\), and to compare this variability against a prescribed \emph{accuracy threshold} \(\tau\) that encodes what we deem to be an acceptable distortion of the posterior. In Sec.~\ref{sec:acc_thresh} we instantiate \(\tau\) using the mean-zeroed variability induced by different physical noise realisations, but the bootstrapping logic itself is independent of that particular choice. It only assumes that we have
(i) an approximately unbiased sampler for the posterior, and
(ii) a way of quantifying distances between posteriors via the divergences \(Y^q_\mathrm{est}(\cdot)\) defined in Eq.~(\ref{setY}).

Operationally, we assume that the additional \emph{downsampling noise} is monotone in \(N_\s\): as \(N_\s\) is reduced, the downsampling-induced variability in the posterior increases (or at least does not decrease). The goal is then to identify a \emph{minimal} value \(N_\s^{\min}\) above which the variability induced by downsampling is always below the accuracy threshold \(\tau\), and beyond which the posterior is effectively independent of further increases in \(N_\s\).

\begin{enumerate}
  \item \textbf{Monotonicity assumption.}
  We assume that, for fixed system and sampler settings, the variability induced purely by downsampling is a non-increasing function of \(N_\s\). Hence there should exist a minimal value \(N_\s^{\min}\) such that, for all \(N_\s \ge N_\s^{\min}\), the downsampling noise is acceptable in the sense that it does not exceed the pre-defined accuracy threshold \(\tau\).

  \item \textbf{Candidate choice \(\hat{N}_\s\) and generation of posteriors.}
  Choose a candidate value \(\hat{N}_\s\) that is expected to be ``sufficiently large''. For this \(\hat{N}_\s\), construct a representative target posterior
  \[
    q(\theta) \equiv p_{\mathbf{k'}}(\theta)
  \]
  corresponding to some fixed sample-selection vector \(\mathbf{k'}\) with \(\hat{N}_\s\) ones (for example using a zero-noise realisation, as described in Sec.~\ref{sec:acc_thresh}). Then generate a set of \(K\) downsampled posteriors
  \(\{p^i_\mathrm{DS+smp}(\theta;\hat{N}_\s)\}\) using different random realisations of \(\mathbf{k}\) with \(\hat{N}_\s\) ones.

  \item \textbf{Acceptance test for \(\hat{N}_\s\).}
  For each of the \(K\) realisations, compute the divergence from the target,
\begin{equation*}\label{eq:YqDSS}
  Y^q_\mathrm{DS+smp}(\hat{N}_\s)
  \equiv
  \Big\{
    \overline{D}_\mathrm{KL}\!\left(
      q \,\Big\|\, p^i_\mathrm{DS+smp}(\hat{N}_\s)
    \right)
  \Big\}_{i=1}^K .
\end{equation*}

  and form the empirical mean \(\mu\big[Y^q_\mathrm{DS+smp}(\hat{N}_\s)\big]\). We then compare this mean divergence with the accuracy threshold \(\tau\):
  \[
    \mu\big[Y^q_\mathrm{DS+smp}(\hat{N}_\s)\big] \;\lesssim\; \tau \, .
  \]
  If this condition is satisfied, then \(\hat{N}_\s\) is deemed \emph{admissible}. If not, \(\hat{N}_\s\) is too small, and we increase \(\hat{N}_\s\) and repeat Steps~2--3. The smallest admissible value is denoted \(N_\s^{\min}\).

  \item \textbf{Posterior stability near \(N_\s^{\min}\).}
  To verify that \(N_\s^{\min}\) lies in a regime where the posterior is stable, retain a representative posterior
  \[
    q(\theta) \equiv p^i_\mathrm{DS+smp}(\theta;N_\s^{\min})
  \]
  as the target distribution. Generate additional downsampled posteriors for nearby values \(N_\s > N_\s^{\min}\) (e.g.\ \(N_\s = 2N_\s^{\min}\)), compute the corresponding divergences \(Y^q_\mathrm{DS+smp}(N_\s)\), and check that
\begin{align*}
  \mu\big[Y^q_\mathrm{DS+smp}(N_\s)\big]
  &\approx
  \mu\big[Y^q_\mathrm{DS+smp}(N_\s^{\min})\big],
  \\
  &\text{for all } N_\s > N_\s^{\min}.
\end{align*}
  This confirms that once \(N_\s\) exceeds \(N_\s^{\min}\), the posterior geometry is effectively independent of the precise choice of \(N_\s\) within the admissible range.

  \item \textbf{Consistency with monotonicity and definition of \(N_\s^\mathrm{opt}\).}
  Finally, examine \(\mu\big[Y^q_\mathrm{DS+smp}(N_\s)\big]\) as a function of \(N_\s\) below and above \(N_\s^{\min}\). If the mean divergence increases (or remains constant) as \(N_\s\) decreases below \(N_\s^{\min}\), the qualitative behaviour assumed in Step~1 is confirmed. One may then choose an ``optimal'' downsampling rate \(N_\s^\mathrm{opt}\ge N_\s^{\min}\), for example the smallest value at which the criteria in Eqs.~(\ref{noise_criterion})--(\ref{avg_entropy}) (Sec.~\ref{sec:acc_thresh}) are satisfied.
\end{enumerate}

In summary, this bootstrapping procedure uses the monotonic dependence of downsampling noise on \(N_\s\) together with an externally defined accuracy threshold \(\tau\) (based, in our case, on mean-zeroed physical noise realisations) to identify a minimal and, subsequently, an optimal downsampling rate.

\subsubsection{Accuracy threshold}\label{sec:acc_thresh}
As discussed in Sec.~\ref{sec:postnoisesources}, different realisations of detector noise induce variability in the posterior that we regard as irreducible in realistic data analyses. We are interested only in changes to the \emph{geometry} of the posterior, not in noise-induced shifts of its mean, so we work with mean-zeroed posteriors when quantifying these effects. Using this setup, we now define an accuracy threshold~\(\tau\) based on the typical geometry variation induced by physical noise.

The downsampling posterior-noise limit is then found as follows.
\begin{enumerate}
    \item Choose some sample-selection vector $\mathbf{k'}$ with $N_\s$ ones.
    \item Produce a set $\{p^i_{\mathbf{k'},\mathrm{phys}}\}$ of $K$ posterior estimates defined by adding different random noise realisations to the (prewhitened, normalised) residuals.
    \item Define a new, mean-zeroed set
    \[
      \left\{\hat{p}^i_{\mathbf{k'},\mathrm{phys}}(\boldsymbol{\theta})\right\}
      =
      \left\{p^i_{\mathbf{k'},\mathrm{phys}}(\boldsymbol{\theta}-\bar{\boldsymbol{\theta}}^i_{\mathbf{k'},\mathrm{phys}})\right\},
    \]
    where $\bar{\boldsymbol{\theta}}^i_{\mathbf{k'},\mathrm{phys}}$ is the vector of posterior means of $p^i_{\mathbf{k'},\mathrm{phys}}$.
\end{enumerate}

Since a vanishing detector-noise realisation, $\mathbf{n}=\boldsymbol{0}$, yields the expected posterior distribution over noise realisations \cite{GPULISA, AvgNoise}, and may also be taken as representative of the set of downsampled posteriors, the zero-noise posterior is the ideal target distribution for comparing downsampling noise to physical noise. It must also be mean-zeroed, so we define the target distribution
\[
  \hat{q}(\boldsymbol{\theta})
  = p_{\mathbf{k'}}(\boldsymbol{\theta}-\bar{\boldsymbol{\theta}}_{\mathbf{k'}}),
\]
where $\bar{\boldsymbol{\theta}}_{\mathbf{k'}}$ is the mean of the zero-noise posterior $p_{\mathbf{k'}}$. With this $\hat{q}$, and the set of $K$ mean-zeroed physical-noise posteriors $\{\hat{p}^i_{\mathbf{k'},\mathrm{phys}}\}$, we form the set of $K$ divergences $\hat{Y}^{\hat{q}}_\mathrm{phys}$ using Eq.~(\ref{setY}). The resulting average posterior ``geometry divergence'' induced by different noise realisations, $\mu[\hat{Y}^{\hat{q}}_\mathrm{phys}]$, is then used to set an acceptance threshold for divergence induced by downsampling.

We define the accuracy threshold
\begin{equation}
  \tau \;\equiv\; \mu\big[\hat{Y}^{\hat{q}}_\mathrm{phys}\big] \, .
\end{equation}
This $\tau$ is then used in Sec.~\ref{sec:bootstrapping} as the reference
level against which downsampling-induced variability is judged.

We now demand the following acceptability criteria for downsampled posteriors: \emph{either}
\begin{subequations}\label{noise_criterion}
\begin{align}
    \mu\big[Y^q_\mathrm{DS+smp}(N_\s)\big] &\lesssim \tau , \label{noise_criteriona} \\
    \mu\big[Y^q_\mathrm{DS+smp}(N_\s)\big] &\approx \mu\big[Y^q_\mathrm{DS+smp}(N'_\s)\big] \, ,\label{noise_criterionb}
\end{align}
\end{subequations}
for all $N_\s, N'_\s > N_\s^\mathrm{opt}$, \emph{or},
\begin{subequations}\label{avg_entropy}
\begin{align}
    \mu\big[Y^q_\mathrm{DS+smp}(\hat{N}_\s)\big] &\approx \tau \, , \\
    \mu\big[Y^q_\mathrm{DS+smp}(N_\s)\big] &\ll \mathcal{H}_q \, , 
\end{align}
\end{subequations}
for all $N_\s > N_\s^\mathrm{opt}$, for some $\Ns^\mathrm{opt}$, where $\mathcal{H}_q$ is the entropy of the target distribution $q$.

Even though the empirical distributions of divergences are non-negative and typically skewed, \(\mu[X]\) remains a convenient summary statistic. Equation~(\ref{noise_criteriona}) ensures that the posterior obtained by downsampling has roughly the same divergence from the target as the average divergence induced by random physical noise realisations. Condition~(\ref{noise_criterionb}) then strongly indicates convergence of the posterior, in the sense that the posterior structure is insensitive to the number of samples used within the admissible range. 

The alternative criteria in (\ref{avg_entropy}) are required when downsampling still affects the posterior geometry as much as physical noise even at the candidate \(\hat{N}_\s\) (this can occur when physical noise mainly shifts the location but does not significantly change the geometry of the posterior). In such cases the average divergences from the target as functions of \(N_\s\) may be very small, and the absolute values in (\ref{noise_criterion}) can differ by orders of magnitude; what matters is that the divergence remains small compared to the intrinsic information scale \(\mathcal{H}_q\). The minimum $N_\s^\mathrm{opt}$ for which the above holds is then taken to be the optimal downsampling rate.

\subsubsection{Summary}\label{sec:criteriasummary}
In this section we have introduced a set of conditions which, taken together,
constitute our acceptance criteria for declaring that a downsampled posterior
has converged to the true posterior at a given $N_\s$, up to an accuracy
comparable to the geometry variation induced by physical detector noise.

\begin{enumerate}
  \item Instantiate the accuracy threshold $\tau$ using physical-noise
  realisations as described in Sec.~\ref{sec:acc_thresh}, i.e.\ set
  $\tau \equiv \mu[\hat{Y}^{\hat{q}}_\mathrm{phys}]$ for the corresponding
  mean-zeroed physical-noise divergences.

  \item Use the bootstrapping procedure of Sec.~\ref{sec:bootstrapping} with
  this accuracy threshold $\tau$ to identify a minimal admissible
  downsampling rate $N_\s^{\min}$ and a corresponding target distribution
  $q \approx p_\mathrm{true}$, chosen as a representative downsampled posterior
  at $N_\s^{\min}$, such that
  \begin{itemize}
    \item the mean divergence from $q$ at $N_\s^{\min}$ satisfies
      $\mu\big[Y^q_\mathrm{DS+smp}(N_\s^{\min})\big] \lesssim \tau$, and
    \item for all $N_\s > N_\s^{\min}$, the mean divergence is stable,
      $\mu\big[Y^q_\mathrm{DS+smp}(N_\s)\big] \approx
       \mu\big[Y^q_\mathrm{DS+smp}(N_\s^{\min})\big]$.
  \end{itemize}

  \item Finally, within the admissible regime $N_\s \ge N_\s^{\min}$, find the
  smallest $N_\s^\mathrm{opt}$ for which either the criteria in
  Eqs.~(\ref{noise_criterion}) or those in Eqs.~(\ref{avg_entropy}) are
  satisfied for all $N_\s > N_\s^\mathrm{opt}$. This $N_\s^\mathrm{opt}$ is
  then taken to be the optimal downsampling rate.
\end{enumerate}

\subsection{Modelling and Analysis Specifics}\label{sec:specifics}

\subsubsection{Structure of analysis}\label{sec:analysis_struct}

We study posteriors across a range of downsampling parameters and signal types, chosen to be representative of envisaged future LISA analyses. These include different downsampling schemes (random/hybrid/cluster), the downsampling method (single factor/FIM preservation), original number of samples in signal $N_\f$, fraction of frequency covered by the signal $\Delta f/f_\mathrm{max}$, where $\Delta f=f_\mathrm{max}-f_\mathrm{min}$, and of course the reduced number of samples $N_\s$. We define the space of injection signal parameters by
\begin{align*}
    N_f&\in\{10^6,10^7,10^8\} \, , \\
    \Delta f/f_\mathrm{max}&\in\{0.9,0.09,0.009\}\, , \\
    N_\s&\in \{16, 23, 32, 45, 64, ..., 8192\} \, .
\end{align*}
That is, $\Ns(n)=\lfloor 2^{n/2}+\tfrac{1}{2}\rfloor$ samples, where $8\leq n \leq 26$ and where $\lfloor x + \tfrac{1}{2}\rfloor$ rounds $x$ to its nearest integer. We will fix $f_\mathrm{max}=0.1 \,\mathrm{Hz}$ for all systems being tested. This parametrisation of the fiducial systems is chosen to probe how the downsampling procedure behaves as a function of generic signal properties (e.g.\ frequency span in the Fourier domain and susceptibility to aliasing or degradation under downsampling).

As we need to quantify the posterior noise for each combination of system parameters, we require a number of posteriors for each combination. We produce 21 posteriors per combination, giving ${21 \choose 2} = 210$ divergences in the sets $X_\mathrm{est}$, and 21 divergences in the sets $Y^q_\mathrm{est}$. Thus, for each downsampling scheme and each method, this equates to $3\times 3\times 22 \times 21 = 4158$ posterior distributions. Since there are also $3\times 2 = 6$ combinations of scheme and method, this would yield $6\times 4158 = 24948$ posteriors in total. To reduce the workload, we therefore split the analysis into two parts. The first part uses a single injection signal to determine the optimal downsampling scheme and method; the second part uses that scheme and method to determine the optimal downsampling rate over the injection-signal parameter space.

\subsubsection{Waveform model}
The parameterised waveform model we use is the inspiral stage of the time-domain waveform of general relativistic binary black holes, as detailed in Ref.\,\cite{IMRPhenomTP}, with the following 8 free parameters: chirp mass ($\mathcal{M}_\mathrm{c}$), (inverted) mass ratio ($q$), effective spin ($\chi_\mathrm{eff}$), distance ($d$), inclination angle ($\theta_\mathrm{JN}$), polarisation angle ($\psi$), coalescence time ($t_\mathrm{c}$), and finally, the coalescence phase ($\phi_\mathrm{c}$), which, however, is marginalised numerically in the likelihood (see Appendix B), leaving 7 free parameters in the final posterior. To convert the waveform parameters into the signal property variables noted in the previous subsection, we can invert the well-known formulas for the inspiral part of BHB waveforms (see, e.g., Ref. \cite{Maggiore})
\begin{align}
    T_\mathrm{obs}&=\tau_{f_\mathrm{min}} - \tau_{f_\mathrm{max}}  \, , \\
    \Nf &= f_\mathrm{Nyq}T_\mathrm{obs} = 2f_\mathrm{max}T_\mathrm{obs} \, , \\
    f(\tau)&=134\,\mathrm{Hz}\left(\frac{1.21 M_\odot}{\mathcal{M}_\mathrm{c}}\right)^{5/8}\left(\frac{1\,\mathrm{s}}{\tau}\right)^{3/8} \, , \label{f_t_eqn}
\end{align}
where $M_\odot$ is the unit of solar mass, $\tau = t_\mathrm{coal}-t$, and $t_\mathrm{coal}$ is the coalescence time, i.e., $\tau$ is the time to coalescence. Inverting to find the required waveform model and signal parameters in terms of $\Nf$ and $\Delta f/f_\mathrm{max}$ (and noting that $f_\mathrm{max}=0.1\, \mathrm{Hz}$ for all systems) we obtain
\begin{align}
    T_\mathrm{obs}&=\frac{\Nf}{2 f_\mathrm{max}} \, , \\
    \tau_{f_\mathrm{max}} &= T_\mathrm{obs}\left(\left(1-\frac{\Delta f}{f_\mathrm{max}}\right)^{-8/3} -1\right)^{-1} \, , \\
    \mathcal{M}_\mathrm{c} &= 1.21\, M_\odot \left(\frac{134 \,\mathrm{Hz}}{f_\mathrm{max}}\right)^{8/5}\left(\frac{1\, \mathrm{s}}{\tau_{f_\mathrm{max}}}\right)^{3/5} \, . \label{Mc_eqn}
\end{align}
As mentioned in the previous subsection, changing the BHB parameters of course modifies the posterior, but is less directly connected to the effects of downsampling. Hence, for all fiducial systems we keep the remaining injection parameters fixed, as listed in Table~\ref{Tab:systems_new}.

\begin{tablehere}
\begin{center}
\small

\begin{tabular}{|| c | c | c | c | c | c | c ||}
 \hline
 $q$ & $\chi_\mathrm{eff}$ & $d$ & $\theta_\mathrm{JN}$ & $\psi$ & $t_\mathrm{c}$ & $\phi_\mathrm{c}$ \\
\hline 
0.8 & 0.32 & 410 Mpc & 0.68 rad & 0.659 rad & 0.0 s & 0.5 rad \\
\hline 
\end{tabular}\caption[BHB test system parameters for assessing the downsampling procedure.]{The remaining BHB test system injection parameters.
\label{Tab:systems_new}}
\end{center}
\end{tablehere}

\subsubsection{Detector model}
In practice, the frequency of the measured waveform will be modulated by LISA's motion around the Sun. This effect enhances sky localisation and introduces sidebands into the waveform. To keep the frequency range small (and thereby reduce the MCS and waveform-evaluation cost for this large suite of posteriors), we omit the sky-position modulation and fix the sky location. We also ignore the tumbling motion and frequency response of LISA, effectively modelling LISA as a LIGO-like detector in a plane with normal vector pointing directly at the source, with stationary noise.

\subsubsection{PSD and MCS}

In order to decrease the sampling rate, we can high-cut/low-pass the PSD at just above the Nyquist rate of the signal, since the frequency-domain inner product \cite{Maggiore} is unaffected if one takes the upper integration limit $\infty\to f_\mathrm{max}$, where
$$f_\mathrm{max}=\max (\supp (\tilde{a}^*\tilde{b} )) \, ,$$
and where $\supp(\cdot)$ stands for the support of the given function. However, what we are really interested in is minimising the MCS, i.e., the number of data points of the residual required to be computed, with minimal affect of the inner product. Simplifying the integral further, we can write
\begin{equation*}
    \langle \tilde{a},\tilde{b} \rangle \equiv \mathfrak{R}\left(4\int_{f_\mathrm{min}}^{f_\mathrm{max}} \d f \frac{\tilde{a}^*(f)\tilde{b}(f)}{\Sn(f)}\right) \, ,
\end{equation*}
where
$$f_\mathrm{min}=\min (\supp (\tilde{a}^*\tilde{b} )) \, ,$$
(note that we use a slightly larger $f_\mathrm{max}$ and slightly smaller $f_\mathrm{min}$ so as to account for the other template signals that will have a non-negligible probability in the posterior).
Therefore, the behaviour of $S_n(f)$ outside the range $f_\mathrm{min} < f < f_\mathrm{max}$ is inconsequential to the inner product, and may be altered in any way convenient for minimising the MCS. We have not attempted a systematic optimisation, but we find that roughly flattening the PSD outside the signal frequency range, by using
\[
  S'_n(f) = \begin{cases}
    S_n(f_\mathrm{min}) & \text{for } f<f_\mathrm{min} \, ,\\
    S_n(f) & \text{for } f_\mathrm{min}<f<f_\mathrm{max} \, ,\\
    S_n(f_\mathrm{max}) & \text{for } f>f_\mathrm{max}\, ,
  \end{cases}
\]
already yields a very low MCS. The MCS for most of our testbed systems (introduced in Section \ref{sec:results}) with unaltered PSD is around 77, and drops to around 5 after this PSD modification; i.e.\ the number of required samples is reduced by roughly a factor of 14. Of all our testbed systems after PSD modification, the highest MCS is 7, so for all subsequent systems we fix $\mathcal{M} = 7$. This allows like-for-like comparison of posterior evaluation times across the parameter space.

Finally, we introduce an overall multiplicative factor in $S'_n(f)$ such that the SNRs of all injected waveforms are equal to 8. We are interested in accurately reproducing generic likelihood functions; at higher SNRs the likelihoods become increasingly Gaussian (lower SNRs exhibit more structure), and in GW data analysis an SNR threshold of $\mathrm{SNR}\gtrsim 8$ is commonly used for detection. Choosing $\mathrm{SNR}=8$ therefore yields posteriors with non-trivial structure while remaining representative of detectable signals.

\subsection{Posterior production}

To obtain posterior estimates, we use the \texttt{Python} Bayesian inference package \texttt{Bilby} \cite{Bilby} with the nested sampler \texttt{Nessai} \cite{Nessai}. The results are sets of samples drawn from the posterior defined by the likelihood function introduced in this paper. The likelihood implementing the downsampling procedures is provided by the \texttt{Python} package \texttt{Dolfen}; see Sec.~\ref{sec:dolfen} for further details.

\subsection{The distances evaluated}\label{sec:results}

\subsubsection{Optimal downsampling scheme and method}\label{sec:part1}

To determine the best downsampling scheme and method, we use one combination of the signal injection parameters described in Sec.~\ref{sec:analysis_struct}, chosen to be least amenable to downsampling: namely, the system that is least well described as slowly evolving (its Fisher information varies most over the signal duration) and whose posterior is most highly structured. For our GW BHB model this corresponds to $\Nf=10^6$ and $\Delta f/f_\mathrm{max}=0.9$.

The target distribution was defined with $\hat{N}_\s=32768$ and satisfies the requirements in Sec.~\ref{sec:criteriasummary}, with $\hat{N}'_\s=8192$ used as the comparison point. For the bootstrapping procedure we use random sampling and the single–factor method based on minimising the Jeffreys' divergence between parameter covariance matrices. Quantitatively, the CMJS divergences arising from sampler noise alone are of order $\sim 5\times 10^{-4}\,\mathrm{bits}$, and including downsampling increases this only slightly (to $\sim 6\times 10^{-4}\,\mathrm{bits}$). The CMKL divergences to the target at $\hat{N}_\s$ and $\hat{N}'_\s$ are $\sim 5$–$7\times 10^{-3}\,\mathrm{bits}$ with standard deviations of order $10^{-2}\,\mathrm{bits}$, so the additional variability from downsampling at these rates is small compared to the intrinsic scatter.

Thus we can see that the first set of conditions as summarised in Sec.~\ref{sec:criteriasummary} are satisfied. The optimal downsampling scheme and method are those that satisfy the second set of conditions in Sec.~\ref{sec:criteriasummary}. The corresponding results are shown in Fig.~\ref{fig:method_conv}; the hybrid scheme combined with the single–factor method (minimising the Jeffreys' divergence of PCMs) appears to yield the best performance.

\begin{figurehere}
   \centering
   \vspace{4pt}
    \includegraphics[scale=0.82]{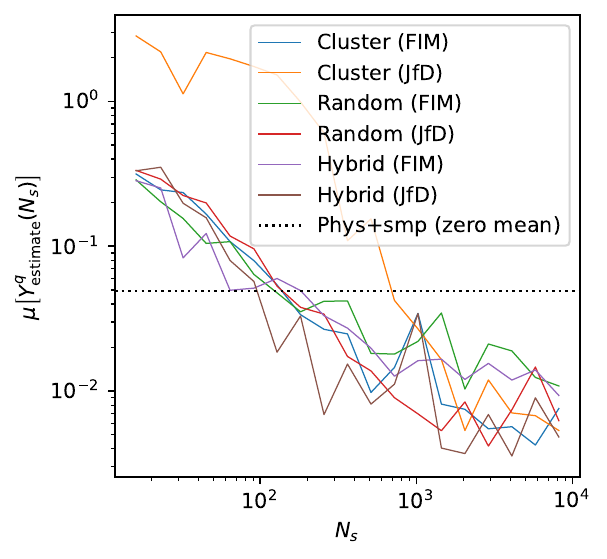}
    \vspace{-22pt}
    \caption[Scheme and method convergence.]{Convergence of the downsampling scheme (data point choice) and method (how to define the new inner product operator). This plot shows the mean of the posterior noise (the mean of the variation in the CMJS distances) induced by downsampling, for sets containing 21 posteriors defined using different random time series sample draws for various downsampling schemes and methods, for posterior sets defined by various $\Ns$. The best performing method and scheme under this metric is that which drops and stays below the (zeroed mean) physical noise (black dashed line) first (from the left). There is roughly equal performance between proposed schemes and methods with the exception of single factor cluster downsampling. The best performing scheme and method, though by a small amount, appears to be hybrid downsampling combined with the single–factor Jeffreys–divergence method.}\label{fig:method_conv}
\end{figurehere}

\subsubsection{Optimal downsampling rate}\label{sec:part2}

\begin{widetext}

\begin{figure}[!h]
   \centering
    \hspace{-1.0cm}\includegraphics[scale=0.75]{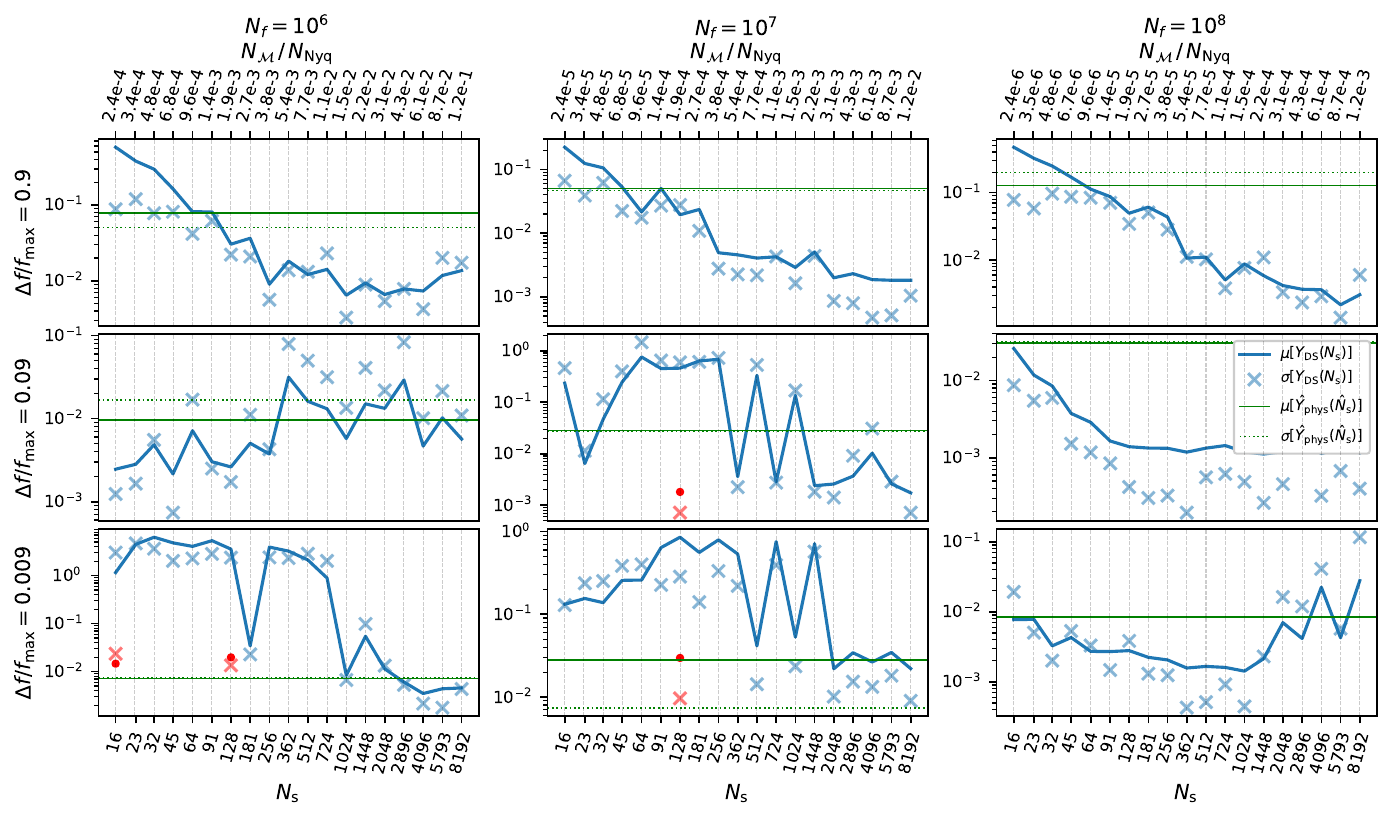}
    \vspace{-5mm}
    \caption[Results of CMJS distances for example systems as function of downsampling rate.]{The means and standard deviations of the sets of \gls{cmjs} divergences, $\overline{D}_\mathrm{JS}$, as functions of $\Ns$, for the systems presented in Section \ref{sec:analysis_struct}. The top axis shows the fraction of the signal's number of Nyquist samples that were required to be computed to evaluate the posteriors. The solid blue line (/blue crosses) shows the CMJS distance mean (/standard deviation) of sets of downsampled posteriors defined using single noise factor, hybrid downsampling, from the target distribution. The solid (/dashed) green lines are the means (/standard deviations) of the posterior noise of zeroed mean distributions defined with physical non-zero noise realisations. The red dots (/crosses) are the means (/standard deviations) of CMJS distances defined using the FIM preservation downsampling method. Convergence criteria is given in Section \ref{sec:convcrit}, and Table \ref{tab:opt_samps} gives $\Ns^\mathrm{opt}$ for each system.} \label{fig:1e6_conv}
\end{figure}
\end{widetext}

We proceed with the optimal downsampling method and scheme found in the previous subsection, and determine its downsampling posterior noise statistics for the set of fiducial systems given in Section \ref{sec:analysis_struct}. The results are shown in Figure \ref{fig:1e6_conv}.

With a strict application of the convergence criteria summarised in Section~\ref{sec:criteriasummary}, we obtain, from Figure~\ref{fig:1e6_conv}, the optimal number of data points as a function of $\Delta f/f_\mathrm{max}$; these values are summarised in Table~\ref{tab:opt_samps}.

Several interesting features are immediately apparent, highlighting both the strengths and the limits of the downsampling method. The observed behaviour seems to arise from a combination of (i) details of our signal model (how data-point values are determined by the parameters) and (ii) the geometry of each posterior (including the influence of the prior distributions).

From Figure~\ref{fig:1e6_conv} and Table~\ref{tab:opt_samps} we see that the behaviour of the downsampling for the $\Delta f/f_\mathrm{max}=0.09$, $\Nf=10^8$ system, and for all $\Delta f/f_\mathrm{max}=0.9$ systems, is as expected. The remaining cases show no obvious pattern. However, on closer inspection of the posteriors and their FIMs, and in light of the signal model and the downsampling procedure, an explanation emerges and an important aspect of downsampling becomes clear.

\def\arraystretch{1.5}
\begin{tablehere}
\small
\begin{center}
\vspace{6pt}
\begin{tabular}{|| c | c | c | c ||} 
\hline

$\Delta f / f_\mathrm{max}$& $\Nf=10^6$  & $\Nf=10^7$  & $\Nf=10^8$ \\
\hline
\hline
$0.9$ & 362 (11.1) [P] & 362 (12.6) [P] & 362 (15.2) [P] \\ \hline
$0.09$ & 16 (12.7) [C2] & 1448 (16.9) [F] & 181 (16.0) [P] \\ \hline
$0.009$ & 1024 (10.0) [F] & 2048 (10.9) [F] & 16 (12.5) [C2] \\ \hline
\end{tabular}
\caption[Table caption]{The optimal number of samples and the entropy (round brackets) of a representative target distribution, for all test systems. These values derive from Figure \ref{fig:1e6_conv} and the convergence criteria given in Section \ref{sec:convcrit}. The cells marked [P] show expected behaviour and comply with convergence criteria in equation (\ref{noise_criterion}). The cells marked [C2] required condition (\ref{avg_entropy}) for acceptance, and those marked [F] show unexpected behaviour, the cause of which appears to be from disproportionate information loss across the parameters. The solution is to use FIM preservation as the downsampling method; this leads to vastly improved results (red dots/crosses in Figure \ref{fig:1e6_conv}), where $\Ns^\mathrm{opt}\leq 128$.
\label{tab:opt_samps}}
\end{center}
\normalsize
\end{tablehere}

Examining the FIMs for posteriors that deviate strongly from the target distributions (and hence exhibit a high level of posterior noise), we find that they typically share the feature that the information on the effective spin parameter is disproportionately large compared with that on the other parameters. In other cases, the relative information content of specific parameter pairs is also anomalously large. Some of these subtler effects are further compounded by the non-white noise covariance matrix, making them difficult to disentangle. Since the FIM is the leading descriptor of the local geometry of the posterior, preserving the Fisher information under downsampling appears to be a natural and promising route to achieving better downsampling rates in these extreme cases.

To test this hypothesis, we computed specific sets of posteriors, 21 in each set, for those systems that converged much later than the `well-behaved' systems, using the FIM preservation method rather than single factor noise reduction, since this method is of course intended to provide accurate covariances. The means and standard deviations of the particular sets we computed are shown as the red dots (means) and crosses (standard deviations) in Figure \ref{fig:1e6_conv}, and indeed very strongly support this hypothesis. Note that for $\Nf=10^6, \Delta f/f_\mathrm{max}=0.009$ we used random rather than hybrid sampling, since \texttt{Dolfen} had difficulty finding a solution for FIM preservation in the hybrid case.

This demonstrates that we chose a fiducial system with which to determine the optimal downsampling scheme and method in Part 1 that was sub-optimal, and which ultimately has led to the sub-optimal choice of downsampling scheme, with which the bulk analysis of determining the optimal number of samples was performed. Clearly using FIM preservation allows $\Ns^\mathrm{opt}\leq 128$ for those systems requiring $\Ns^\mathrm{opt}\geq 1024$ under single factor downsampling. By replacing the values in the cells of Table \ref{tab:opt_samps} marked [F] with the new values, $\Ns=128$, we can still confidently conclude that using FIM preservation downsampling, the optimal downsampling number of samples for our range of tested systems is at most $\Ns^\mathrm{opt}=362$. Most often, one requires far fewer samples than this, however, a general method of computing $\Ns^\mathrm{opt}$ for any given system (without producing and studying many posteriors as we have done) from, say, the Fisher matrix, presently does not appear to be possible.

\section{Speed comparison and further remarks}\label{sec:evaltimes}

\subsection{Posterior estimate evaluation times}\label{sec:PE_times}

The average PE convergence time for each of the $\Delta f/f_\mathrm{max}=0.9$ systems is plotted in Figure \ref{fig:convergencetimes}. As to be expected, the evaluation time drops as the number of data points decreases. However, convergence time reaches a minimum of a few hours at around $\Ns\approx 256$, and begins to rise as $\Ns$ continues to decrease. This is because the posterior structure begins to qualitatively break down, as extra modes are introduced and the general posterior structure becomes highly irregular, since the low number of data points cannot accurately encode the original posterior. The sampler therefore has significantly more difficulty in finding the isoprobability contours, and thus requires more time to converge, despite only computing a small number of data point values to evaluate the likelihood.

Increasing $\Ns$ beyond $\Ns^\mathrm{opt}$ only increases the amount of time-series data point computations required to evaluate the likelihood, without affecting the likelihood (by definition), thus the posterior evaluation time is expected to increase roughly linearly with increasing $\Ns$. We see in Fig.~\ref{fig:convergencetimes} that this is indeed the case; a line of best fit is drawn through those average times for which $\Ns\geq \Ns^\mathrm{opt}$. We model the mean PE time as linear in the number of (effective) samples:
\begin{equation}\label{linearPEtimemodel}
    T(N) = aN + c \, ,
\end{equation}
where $T$ is the average evaluation time, $N$ is the number of retained time–domain samples, $a$ is the slope and $c$ is a constant time offset (attributed to sampler overheads: initialisation, Fisher–matrix computation, normalising–flow training in \texttt{Nessai}, etc.).

Using Eq.~(\ref{MCSsamples}), the number of residual points that must be computed for a given downsampling choice is $N_\mathcal{M}(\Ns)\approx (2\mathcal{M}+1)\Ns$. Rewriting $T$ as a function of $N_\mathcal{M}$ gives
\begin{align}\label{linearPEtimemodel2}
    T(\Ns) &\approx a\Ns + c
            = a\frac{N_\mathcal{M}(\Ns)}{2\mathcal{M}+1} + c
            \equiv b\,N_\mathcal{M}(\Ns) + c \, , \numberthis
\end{align}
so that $b = a(2\mathcal{M}+1)^{-1}$. Here $b$ is the effective slope when time is expressed as a function of the number of computed residuals $N_\mathcal{M}$ rather than the retained Nyquist samples $\Ns$.

The linear model of posterior evaluation time can be used to estimate the cost
of evaluating the likelihood on all samples ($N_\mathcal{M}=N_\mathrm{Nyq}$),
and hence to gauge the time saved by downsampling. Simulated LISA analyses are
often performed in the frequency domain, where the inner product is especially
cheap because the data samples are independent. A direct comparison of
time-domain and frequency-domain evaluation times is therefore only approximate:
the number of numerical operations required to compute templates differs
slightly between the two, and our time-domain scheme includes a few additional
element-wise products (Sec.~\ref{timedomaincorr}). However, these extra
operations are negligible compared to the cost of evaluating realistic
waveforms $h(t)$, particularly for more detailed signal models with many
terms. With these caveats in mind, Table~\ref{Tab:convergencetimes} reports the
fitted linear parameters for the time-domain evaluation cost and compares the
expected evaluation time of the original posterior with that of the
downsampled posterior.

Note that in Table \ref{Tab:convergencetimes} we have
\begin{equation}\label{rateminusoverhead}
\Delta T \equiv T - c \, ,
\end{equation}
so that, as well as comparing the estimated evaluation times of the original and downsampled posteriors (minus overheads) which of course is of primary interest, we can also see that the quotient is simply given by
\begin{equation}\label{speedratio}
    \frac{\Delta T(N_1)}{\Delta T(N_2)} = \frac{aN_1}{aN_2} = \frac{N_1}{N_2} \, ,
\end{equation}
independent of the slope. The fraction of the estimated time of convergence of the original posterior (minus overheads) required for convergence of approximation by downsampling is $N_\mathcal{M}(\Ns^\mathrm{opt})/\Nf$.

\begin{figurehere}
   \centering
    \vspace{4pt}
    \includegraphics[scale=0.85]{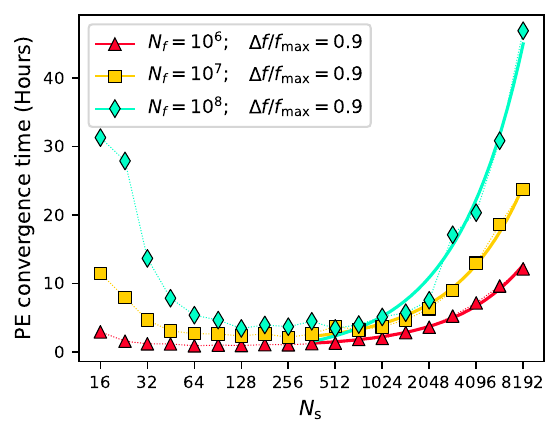}
    \vspace{-20pt}
    \caption[Parameter estimation convergence times]{The average convergence times for selected \gls{bhb} systems. The relationship is strongly linear after $\Ns\gtrsim 362$ since the posterior has stabilised and takes its simplest form; for $\Ns\lesssim 128$, posterior estimation time increases as the number of samples decreases, since the structure of the posterior begins to break down, becoming highly irregular and multi-modal, and more difficult for the sampler to explore.}
    \label{fig:convergencetimes}
\end{figurehere}

\vspace{4pt}

\begin{widetext}

\def\arraystretch{1.2}
\begin{adjustwidth}{-0.0cm}{}
\begin{table}
\small
\begin{center}
\begin{tabular}{|| c | c | c | c | c | c | c | c ||} 
 \hline
$\Nf$ & $\Delta f/f_\mathrm{max}$ & $a$ & $b$ & $c$ & $\Delta T_{\Nf}$ & $\Delta T_{\Ns^\mathrm{opt}}$ & $\Delta T_{\Nf}/\Delta T_{\Ns^\mathrm{opt}}$ \\ [0.5ex] 
 \hline\hline
$10^{6}$ & 0.009 & $2.28 \times 10^{-3}$ & $1.52 \times 10^{-4}$ & 3.10 & $1.5 \times 10^{2}$ & 0.82 & $1.84 \times 10^{2}$ \\ \hline 
$10^{6}$ & 0.09 & $6.18 \times 10^{-3}$ & $4.12 \times 10^{-4}$ & 4.38 & $4.1 \times 10^{2}$ & 2.24 & $1.84 \times 10^{2}$ \\ \hline 
$10^{6}$ & 0.9 & $1.40 \times 10^{-3}$ & $9.31 \times 10^{-5}$ & 1.05 & $9.3 \times 10^{1}$ & 0.51 & $1.84 \times 10^{2}$ \\ \hline 
$10^{7}$ & 0.009 & $4.69 \times 10^{-3}$ & $3.13 \times 10^{-4}$ & 5.50 & $3.1 \times 10^{3}$ & 1.70 & $1.84 \times 10^{3}$ \\ \hline 
$10^{7}$ & 0.09 & $4.30 \times 10^{-3}$ & $2.86 \times 10^{-4}$ & 20.46 & $2.9 \times 10^{3}$ & 1.56 & $1.84 \times 10^{3}$ \\ \hline 
$10^{7}$ & 0.9 & $2.40 \times 10^{-3}$ & $1.60 \times 10^{-4}$ & 3.21 &  $1.6 \times 10^{3}$ & 0.87 & $1.84 \times 10^{3}$ \\ \hline 
$10^{8}$ & 0.009 & $6.43 \times 10^{-3}$ & $4.29 \times 10^{-4}$ & 5.50 &  $4.3 \times 10^{4}$ & 2.33 & $1.84 \times 10^{4}$ \\ \hline 
$10^{8}$ & 0.09 & $4.51 \times 10^{-3}$ & $3.01 \times 10^{-4}$ & 15.46 & $3.0 \times 10^{4}$ & 1.63 & $1.84 \times 10^{4}$ \\ \hline 
$10^{8}$ & 0.9 & $3.99 \times 10^{-3}$ & $2.66 \times 10^{-4}$ & 7.14 & $2.7 \times 10^{4}$ & 1.45 & $1.84 \times 10^{4}$ \\ \hline 
\end{tabular}\caption[Best fit model of posterior evaluation times]{Details of the linear best fit model of posterior evaluation times. The parameters $a, b, c$, are from Equations (\ref{linearPEtimemodel}) \& (\ref{linearPEtimemodel2}). Also shown is the estimated original posterior evaluation time \emph{minus the overhead time}, $c$, (hours), $\Delta T_{\Nf}=\Delta T(N=\Nf)$ as defined in Equation (\ref{rateminusoverhead}), the average downsampled posterior (using $\Ns^\mathrm{opt}\!=\!362$) evaluation time minus the overhead time (hours), $\Delta T_{\Ns^\mathrm{opt}}=\Delta T(N=362)$ and finally the ratio of these times (this is the approximate time saving factor of the posterior evaluation time, after disregarding the constant overheads). Recall these values are computed by setting $\mathcal{M}=7$ for all systems, the highest $\mathcal{M}$ of all systems. Most systems had $\mathcal{M}=4$, however, where $\Delta T_{\Ns}$ would be $\tfrac{3}{5}$ of the value given here, had we not manually set $\mathcal{M}=7$.
\label{Tab:convergencetimes}}
\end{center}
\end{table}
\end{adjustwidth}

\end{widetext}
\subsection{Discussion}

The rightmost column of Table~\ref{Tab:convergencetimes} shows that the reduction in convergence time can already be substantial for the test systems considered here. The quoted times correspond to data lengths down to $\Nf \sim 10^6$, which is conservative compared to the longest LISA inspirals in a 10-year mission, which can reach $\Nf \sim 10^9$. The tabulated speed-ups are computed using $\Ns = 362$ and $\mathcal{M} = 7$. For some systems, however, convergence was achieved with as few as $16$ samples and a true MCS of $\mathcal{M}=4$. Extrapolate to $\Nf = 10^9$ using these values, we obtain $\Delta T_{\Nf}/\Delta T_{\Ns^\mathrm{opt}} \approx 7\times 10^6$, i.e.\ likelihood evaluations could, in such cases, be up to $\sim 7\times 10^6$ times faster than for the original datasets.

In practice, most LISA signals will have fewer Nyquist data points and many will satisfy $\mathcal{M}>4$. Moreover, since the optimal downsampling rate is not known \emph{a priori}, one typically adopts a  somewhat larger $\Ns$ to avoid misrepresenting the true posterior (recommended usage of \texttt{Dolfen} in this regard is discussed in Sec.~\ref{sec:rec_usage}). Even so, for a 4-year LISA mission, speed-ups of order $10^4$ and above are often realistic.

We have so far focused on $n$-dimensional posteriors that can be regarded as slices of an $(n+1)$-dimensional posterior. In an $(n+1)$-dimensional \gls{pe} problem, the likelihood values on a given $n$-dimensional slice are unchanged by also evaluating the likelihood on neighbouring slices. Since no slice is special, one can expect neighbouring slices to be approximated with similar accuracy, allowing an accurate higher-dimensional posterior to be built up from the same downsampled subset of data points. In this sense, downsampling is not expected to depend strongly on the number of free model parameters, although accuracy should always be checked for very high dimensionality. Informal tests on 17-dimensional posteriors using $\Ns=500$ (the default \texttt{Dolfen} setting) have shown extremely
consistent results, indicating excellent accuracy up to at least 17 parameters.

We have proposed and tested several downsampling methods and schemes. In Sec.~\ref{sec:part1} we initially found that hybrid single-factor downsampling was marginally more efficient, and we therefore used this method to produce large sets of posteriors in Sec.~\ref{sec:PE_times}. Subsequent investigation of the surprising behaviour seen in Fig.~\ref{fig:1e6_conv} and Table~\ref{tab:opt_samps} led us to identify model-dependent effects that can be mitigated by using the FIM-preservation method. In hindsight, Fig.~\ref{fig:method_conv} already contained a clue: all methods and schemes broadly agree, except for the cluster scheme, which performs poorly with single-factor downsampling but improves dramatically when FIM preservation is used. The red markers in Fig.~\ref{fig:method_conv} show that, at $\Ns=128$, the FIM-preservation statistics are already comparable to those of the converged posteriors obtained with single-factor noise reduction ($\Ns\gtrsim 1448$).

Our results demonstrate that downsampling is a viable way to reproduce the likelihood function cheaply and accurately in simulated environments. However, we have not found any reliable method of predicting $\Ns^\mathrm{opt}$ for a given \gls{pe} problem from simpler diagnostic quantities (such as the FIM) within the standard framework. For readers familiar with machine learning, it is useful to view the downsampled data space as analogous to a \emph{latent space}: a reduced representation that stands in for the full high-dimensional data. In machine learning, the map from latent variables to data is learned; here, downsampling instead exploits restrictions of a map already provided by the signal model, refined by the optimisation procedures described in this paper. Equivalently, one may regard the signal and noise model evaluated at the retained data points as an encoding of the posterior geometry. Without prior knowledge of the posterior, however, we do not know in advance what should be encoded, and downsampling is fundamentally limited by this; perfect accuracy can never be guaranteed. Nevertheless, using FIM preservation rather than minimising Jeffreys divergences of the PCMs yields substantially improved and highly accurate results for our example systems.

For most applications, the default \texttt{Dolfen} settings will likely produce sufficiently accurate posteriors and should be more than adequate for exploratory or informal investigations. When a high level of certainty is required, we recommend explicitly checking convergence by generating posteriors at several values of $\Ns$ and verifying that the resulting posteriors are effectively independent of $\Ns$ within the admissible range.

Although we have seen dramatic improvements in \gls{pe} evaluation times, many realistic signal models will be more intricate than our standard BHB inspiral signal, with likelihood functions that are more expensive to compute. Approximation techniques such as the one presented here, and others like \gls{roq}, will be essential tools for experimentation with \gls{lisa}-like signals whose full likelihoods are prohibitively costly to evaluate. A natural direction for future work is to investigate how these approaches can be combined to provide an even more rapid feedback arena for studying \gls{bhb} signals and their waveform modifications in the LISA context.

Finally, in applications that aim to quantify systematic modelling bias, one may precompute the Fisher information matrix for the injected signal (generated with the ``true'' physics) and use it to define the downsampled inner product. In this setup, the downsampled data preserve the (Fisher) information content of the original signal about the true parameters, while the same inner product is then used to evaluate the likelihood under a mismatched template model. This is a natural strategy for bias studies, but the resulting likelihood still needs to be validated for the chosen template. The key advantage is that the approximation scheme is held fixed, so any observed bias in parameter recovery can be attributed to the physics mismatch itself, rather than to changes in the numerical representation of the data.

\section{The \texttt{Dolfen} package}\label{sec:dolfen}

The \ul{do}wnsampling \ul{l}ikelihood \ul{f}unction \ul{e}stimatio\ul{n} package \texttt{Dolfen}, performs all variants of the downsampling procedure that have been derived and discussed in this article. \texttt{Dolfen} is a \texttt{pip} installable \texttt{Python} package that can easily be used in conjunction with \texttt{Bilby} \cite{Bilby} as per the examples provided on \texttt{Dolfen}'s documentation pages (\texttt{https://dolfen.readthedocs.io}) and \texttt{github} pages (\texttt{https://github.com/jethrolinley/dolfen}).

\subsection{Future developments}
Recall from Section \ref{preserveFIM} that numerical issues often arise in exact FIM preservation. Approximate FIM preservation is far easier to solve, and \texttt{Dolfen} appears to succeed most often in implementing this technique. In a subsequent version of \texttt{Dolfen}, we intend to include another FIM preservation method; rather than find an individual sample reweighting solution for (approximate) FIM preservation in a given basis (see equation (\ref{weightfunc})), we shall use a (pre-existing) numerical optimisation method to find individual data point reweightings (strictly positive) that minimises divergence between the FIMs.

Real LISA streams will contain short veto windows (telemetry drop-outs, thruster events, etc.). In the present release a user can already approximate such data by (i) cutting the timeline into $M$ contiguous, gap-free segments, (ii) running \texttt{dolfen.log\_likelihood} on each segment independently, and (iii) summing the log-likelihoods. Because the segments are disjoint, this is exactly the ``\emph{cluster-sampling}'' factorisation introduced in Section \ref{dsschemes}, with clusters chosen by the user rather than by the current fixed method. The next version of \texttt{Dolfen} will automate this procedure, where the user can list time intervals to be excluded.

\subsection{Recommended usage}\label{sec:rec_usage}

To assert that an approximate posterior is \emph{close} to the true posterior, one would in principle need access to the true posterior itself in order to compare them directly. Of course, if the true posterior were already known, no approximation would be required. Downsampling instead makes an informed attempt to optimise the encoding of information contained in a randomly selected subset of data points, yielding a lossy compression of the likelihood. As with any likelihood approximation, its accuracy must be assessed in context.

We therefore recommend treating downsampling as a controlled approximation and checking convergence with respect to the number of retained samples, $\Ns$, and the resulting MCS, $\mathcal{M}$. In \texttt{Dolfen}, both $\Ns$ and $\mathcal{M}$ can be explicitly set, overriding the automatically computed MCS if desired. A practical workflow is:

\begin{enumerate}
  \item \textbf{Exploratory run.} Start with a deliberately small choice, e.g.\ $\Ns = 100$ and $\mathcal{M} = 1$. The resulting likelihood should not be expected to be highly accurate, but it is typically sufficient to explore the rough likelihood geometry, tune priors, and adjust sampler settings.
  \item \textbf{Default (high-accuracy) runs.} Once a sensible configuration has been identified, run \texttt{Dolfen} with its default settings to obtain a \emph{likely accurate} likelihood.
  \item \textbf{Convergence check.} Generate several (e.g.\ three) posterior samples using different random selections of data points (with zero noise realisations) and different values of $\Ns$ (for example, $\Ns = 300, 600, 900$). If the resulting posteriors are mutually consistent, this provides strong evidence that the approximation has converged. If not, increase $\Ns$ and repeat.
\end{enumerate}

In practice, it is highly implausible that several genuinely poor and mutually distinct likelihood approximations would, by chance, yield closely similar posteriors. Consequently, stability of the posterior under changes in $\Ns$ and random subsampling provides a practical and robust diagnostic of approximation accuracy.

\section{Acknowledgements}
The author is grateful to Graham Woan and John Veitch for many helpful discussions, and to Michael Williams, for help in operating his sampler $\texttt{Nessai}$. This work was partly funded by STFC grant ST/S505390/1. We are grateful for computational resources provided by Cardiff University, and funded by an STFC grant supporting UK Involvement in the Operation of Advanced LIGO. \emph{Software:} \texttt{Dolfen} is implemented in \texttt{Python} and uses \texttt{NumPy} \cite{numpy} and \texttt{SciPy} \cite{scipy}.

\end{multicols}

\printbibliography 

@misc{cornish2013fastfishermatriceslazy,
      title={Fast Fisher Matrices and Lazy Likelihoods}, 
      author={Neil J. Cornish},
      year={2013},
      eprint={1007.4820},
      archivePrefix={arXiv},
      primaryClass={gr-qc},
      url={https://arxiv.org/abs/1007.4820}, 
}

@misc{capuano2025,
      title={Systematic bias in LISA ringdown analysis due to waveform inaccuracy}, 
      author={Lodovico Capuano and Massimo Vaglio and Rohit S. Chandramouli and Chantal L Pitte and Adrien Kuntz and Enrico Barausse},
      year={2025},
      eprint={2506.21181},
      archivePrefix={arXiv},
      primaryClass={gr-qc},
      url={https://arxiv.org/abs/2506.21181}, 
}

@article{Kapil2024,
   title={Systematic bias from waveform modeling for binary black hole populations in next-generation gravitational wave detectors},
   volume={109},
   ISSN={2470-0029},
   url={http://dx.doi.org/10.1103/PhysRevD.109.104043},
   DOI={10.1103/physrevd.109.104043},
   number={10},
   journal={Physical Review D},
   publisher={American Physical Society (APS)},
   author={Kapil, Veome and Reali, Luca and Cotesta, Roberto and Berti, Emanuele},
   year={2024},
   month=may }

@ARTICLE{speri,
AUTHOR={Speri, Lorenzo and Katz, Michael L. and Chua, Alvin J. K. and Hughes, Scott A. and Warburton, Niels and Thompson, Jonathan E. and Chapman-Bird, Christian E. A. and Gair, Jonathan R.},
TITLE={Fast and Fourier: extreme mass ratio inspiral waveforms in the frequency domain},
JOURNAL={Frontiers in Applied Mathematics and Statistics},      
VOLUME={9},           
YEAR={2024},      
URL={https://www.frontiersin.org/articles/10.3389/fams.2023.1266739},       
DOI={10.3389/fams.2023.1266739},      
ISSN={2297-4687},   
}

@article{numpy,
  author={van der Walt, Stefan and Colbert, S. Chris and Varoquaux, Gael},
  journal={Computing in Science \& Engineering}, 
  title={The NumPy Array: A Structure for Efficient Numerical Computation}, 
  year={2011},
  volume={13},
  number={2},
  pages={22-30},
  doi={10.1109/MCSE.2011.37}}

@article{scipy,
   title={SciPy 1.0: fundamental algorithms for scientific computing in Python},
   volume={17},
   ISSN={1548-7105},
   url={http://dx.doi.org/10.1038/s41592-019-0686-2},
   DOI={10.1038/s41592-019-0686-2},
   number={3},
   journal={Nature Methods},
   publisher={Springer Science and Business Media LLC},
   author={Virtanen, Pauli and Gommers, Ralf and Oliphant, Travis E. and Haberland, Matt},
   year={2020},
   month=feb, pages={261–272} }

@article{Cornish_2020,
   title={Time-frequency analysis of gravitational wave data},
   volume={102},
   ISSN={2470-0029},
   url={http://dx.doi.org/10.1103/PhysRevD.102.124038},
   DOI={10.1103/physrevd.102.124038},
   number={12},
   journal={Physical Review D},
   publisher={American Physical Society (APS)},
   author={Cornish, Neil J.},
   year={2020},
   month=dec }

@article{TDI,
    author = "Tinto, Massimo and Dhurandhar, Sanjeev V.",
    title = "{Time-delay interferometry}",
    doi = "10.1007/s41114-020-00029-6",
    journal = "Living Rev. Rel.",
    volume = "24",
    number = "1",
    pages = "1",
    year = "2021"
}

@article{IMRPhenomTP,
   title={Phenomenological time domain model for dominant quadrupole gravitational wave signal of coalescing binary black holes},
   volume={103},
   ISSN={2470-0029},
   url={http://dx.doi.org/10.1103/PhysRevD.103.124060},
   DOI={10.1103/physrevd.103.124060},
   number={12},
   journal={Physical Review D},
   publisher={American Physical Society (APS)},
   author={Estellés, Héctor and Ramos-Buades, Antoni and Husa, Sascha and García-Quirós, Cecilio and Colleoni, Marta and Haegel, Leïla and Jaume, Rafel},
   year={2021},
   month={ 06}
}

@article{FisherDef,
  title={Computation of the Fisher information matrix for time series models},
  author={Andr{\'e} Klein and Guy M{\'e}lard},
  journal={Journal of Computational and Applied Mathematics},
  year={1995},
  volume={64},
  pages={57-68}
}

@book{Bretthorst,
  author = {Bretthorst, G Larry},
  publisher = {Springer-Verlag Berlin Heidelberg},
  title = {Bayesian spectrum Analysis and parameter estimation},
  url = {http://bayes.wustl.edu/glb/book.pdf},
  year = 1988
}

@article{InadequateFIM,
   title={Inadequacies of the Fisher information matrix in gravitational-wave parameter estimation},
   volume={88},
   ISSN={1550-2368},
   url={http://dx.doi.org/10.1103/PhysRevD.88.084013},
   DOI={10.1103/physrevd.88.084013},
   number={8},
   journal={Physical Review D},
   publisher={American Physical Society (APS)},
   author={Rodriguez, Carl L. and Farr, Benjamin and Farr, Will M. and Mandel, Ilya},
   year={2013},
   month={ 10}
}

@article{FisherDef2,
   title={Generalisations of Fisher Matrices},
   volume={18},
   ISSN={1099-4300},
   url={http://dx.doi.org/10.3390/e18060236},
   DOI={10.3390/e18060236},
   number={6},
   journal={Entropy},
   publisher={MDPI AG},
   author={Heavens, Alan},
   year={2016},
   month={ 06},
   pages={236}
}

@book{Maggiore,
  title={Gravitational Waves: Volume 1: Theory and Experiments},
  author={Maggiore, M.},
  isbn={9780198570745},
  lccn={2008270556},
  series={Gravitational Waves},
  url={https://books.google.co.uk/books?id=AqVpQgAACAAJ},
  year={2008},
  publisher={OUP Oxford}
}

@article{chirp,
   title={Gravitational Waves from Merging Compact Binaries},
   volume={47},
   ISSN={1545-4282},
   url={http://dx.doi.org/10.1146/annurev-astro-082708-101711},
   DOI={10.1146/annurev-astro-082708-101711},
   number={1},
   journal={Annual Review of Astronomy and Astrophysics},
   publisher={Annual Reviews},
   author={Hughes, Scott A.},
   year={2009},
   month={ 09},
   pages={107–157}
}

@misc{LISAwhitepaper,
      title={The Laser Interferometer Space Antenna: Unveiling the Millihertz Gravitational Wave Sky}, 
      author={John Baker and Jillian Bellovary and Peter L. Bender and Emanuele Berti and Robert Caldwell and Jordan Camp and John W. Conklin and Neil Cornish and Curt Cutler and Ryan DeRosa and Michael Eracleous and Elizabeth C. Ferrara and Samuel Francis and Martin Hewitson and Kelly Holley-Bockelmann and Ann Hornschemeier and Craig Hogan and Brittany Kamai and Bernard J. Kelly and Joey Shapiro Key and Shane L. Larson and Jeff Livas and Sridhar Manthripragada and Kirk McKenzie and Sean T. McWilliams and Guido Mueller and Priyamvada Natarajan and Kenji Numata and Norman Rioux and Shannon R. Sankar and Jeremy Schnittman and David Shoemaker and Deirdre Shoemaker and Jacob Slutsky and Robert Spero and Robin Stebbins and Ira Thorpe and Michele Vallisneri and Brent Ware and Peter Wass and Anthony Yu and John Ziemer},
      year={2019},
      eprint={1907.06482},
      archivePrefix={arXiv},
      primaryClass={astro-ph.IM}
}

@article{GrimmHarms,
   title={Multiband gravitational-wave parameter estimation: A study of future detectors},
   volume={102},
   ISSN={2470-0029},
   url={http://dx.doi.org/10.1103/PhysRevD.102.022007},
   DOI={10.1103/physrevd.102.022007},
   number={2},
   journal={Physical Review D},
   publisher={American Physical Society (APS)},
   author={Grimm, Stefan and Harms, Jan},
   year={2020},
   month={ 07}
}

@article{MBinGWTC1,
   title={Multiband observation of LIGO/Virgo binary black hole mergers in the gravitational-wave transient catalog GWTC-1},
   volume={496},
   ISSN={1365-2966},
   url={http://dx.doi.org/10.1093/mnras/staa1512},
   DOI={10.1093/mnras/staa1512},
   number={1},
   journal={Monthly Notices of the Royal Astronomical Society},
   publisher={Oxford University Press (OUP)},
   author={Liu, Chang and Shao, Lijing and Zhao, Junjie and Gao, Yong},
   year={2020},
   month={ 06},
   pages={182–196}
}

@article{FundPhyswGWs,
   title={Probing fundamental physics with gravitational waves: The next generation},
   volume={103},
   ISSN={2470-0029},
   url={http://dx.doi.org/10.1103/PhysRevD.103.044024},
   DOI={10.1103/physrevd.103.044024},
   number={4},
   journal={Physical Review D},
   publisher={American Physical Society (APS)},
   author={Perkins, Scott E. and Yunes, Nicolás and Berti, Emanuele},
   year={2021},
   month={ 02}
}

@article{ExpndLISAfrmGrnd,
   title={Expanding the LISA Horizon from the Ground},
   volume={121},
   ISSN={1079-7114},
   url={http://dx.doi.org/10.1103/PhysRevLett.121.251102},
   DOI={10.1103/physrevlett.121.251102},
   number={25},
   journal={Physical Review Letters},
   publisher={American Physical Society (APS)},
   author={Wong, Kaze W. K. and Kovetz, Ely D. and Cutler, Curt and Berti, Emanuele},
   year={2018},
   month={ 12}
}

@article{LISAEMRIsFisher,
   title={Using LISA extreme-mass-ratio inspiral sources to test off-Kerr deviations in the geometry of massive black holes},
   volume={75},
   ISSN={1550-2368},
   url={http://dx.doi.org/10.1103/PhysRevD.75.042003},
   DOI={10.1103/physrevd.75.042003},
   number={4},
   journal={Physical Review D},
   publisher={American Physical Society (APS)},
   author={Barack, Leor and Cutler, Curt},
   year={2007},
   month={ 02}
}

@article{TransOrbRes,
   title={Assessing the impact of transient orbital resonances},
   volume={103},
   ISSN={2470-0029},
   url={http://dx.doi.org/10.1103/PhysRevD.103.124032},
   DOI={10.1103/physrevd.103.124032},
   number={12},
   journal={Physical Review D},
   publisher={American Physical Society (APS)},
   author={Speri, Lorenzo and Gair, Jonathan R.},
   year={2021},
   month={ 06}
}

@article{UseAbuseFisher,
   title={Use and abuse of the Fisher information matrix in the assessment of gravitational-wave parameter-estimation prospects},
   volume={77},
   ISSN={1550-2368},
   url={http://dx.doi.org/10.1103/PhysRevD.77.042001},
   DOI={10.1103/physrevd.77.042001},
   number={4},
   journal={Physical Review D},
   publisher={American Physical Society (APS)},
   author={Vallisneri, Michele},
   year={2008},
   month={ 02}
}

@article{ROQ,
   title={Fast and accurate inference on gravitational waves from precessing compact binaries},
   volume={94},
   ISSN={2470-0029},
   url={http://dx.doi.org/10.1103/PhysRevD.94.044031},
   DOI={10.1103/physrevd.94.044031},
   number={4},
   journal={Physical Review D},
   publisher={American Physical Society (APS)},
   author={Smith, Rory and Field, Scott E. and Blackburn, Kent and Haster, Carl-Johan and Pürrer, Michael and Raymond, Vivien and Schmidt, Patricia},
   year={2016},
   month={ 08}
}

@misc{RelativeBinning,
      title={Relative Binning and Fast Likelihood Evaluation for Gravitational Wave Parameter Estimation}, 
      author={Barak Zackay and Liang Dai and Tejaswi Venumadhav},
      year={2018},
      eprint={1806.08792},
      archivePrefix={arXiv},
      primaryClass={astro-ph.IM}
}

@article{TemplateInterpolation,
   title={Accelerating gravitational wave parameter estimation with multi-band template interpolation},
   volume={34},
   ISSN={1361-6382},
   url={http://dx.doi.org/10.1088/1361-6382/aa6d44},
   DOI={10.1088/1361-6382/aa6d44},
   number={11},
   journal={Classical and Quantum Gravity},
   publisher={IOP Publishing},
   author={Vinciguerra, Serena and Veitch, John and Mandel, Ilya},
   year={2017},
   month={ 05},
   pages={115006}
}

@article{AdaptFreqRes,
   title={Accelerating parameter estimation of gravitational waves from compact binary coalescence using adaptive frequency resolutions},
   volume={104},
   ISSN={2470-0029},
   url={http://dx.doi.org/10.1103/PhysRevD.104.044062},
   DOI={10.1103/physrevd.104.044062},
   number={4},
   journal={Physical Review D},
   publisher={American Physical Society (APS)},
   author={Morisaki, Soichiro},
   year={2021},
   month={ 08}
}

@article{emcee,
	doi = {10.1086/670067},
 	url = {https://doi.org/10.1086/670067},
 	year = 2013,
	month = {03},
 	publisher = {{IOP} Publishing},
 	volume = {125},
 	number = {925},
 	pages = {306--312},
 	author = {Daniel Foreman-Mackey and David W. Hogg and Dustin Lang and Jonathan Goodman},
 	title = {emcee: The {MCMC} Hammer},
 	journal = {Publications of the Astronomical Society of the Pacific}
}

@ARTICLE{Dynesty,
       author = {{Speagle}, Joshua S.},
        title = "{DYNESTY: a dynamic nested sampling package for estimating Bayesian posteriors and evidences}",
      journal = {mnras},
         year = 2020,
        month = apr,
       volume = {493},
       number = {3},
        pages = {3132-3158},
          doi = {10.1093/mnras/staa278},
archivePrefix = {arXiv},
       eprint = {1904.02180},
 primaryClass = {astro-ph.IM},
       adsurl = {https://ui.adsabs.harvard.edu/abs/2020MNRAS.493.3132S}
}

@MISC{CPnest,
       author = {{Del Pozzo}, Walter and {Veitch}, John},
        title = "{CPNest: Parallel nested sampling}",
 howpublished = {Astrophysics Source Code Library, record ascl:2205.021},
         year = 2022,
        month = may,
          eid = {ascl:2205.021},
        pages = {ascl:2205.021},
archivePrefix = {ascl},
       eprint = {2205.021},
       adsurl = {https://ui.adsabs.harvard.edu/abs/2022ascl.soft05021D}
}

@article{Nessai,
   title={Nested sampling with normalizing flows for gravitational-wave inference},
   volume={103},
   ISSN={2470-0029},
   url={http://dx.doi.org/10.1103/PhysRevD.103.103006},
   DOI={10.1103/physrevd.103.103006},
   number={10},
   journal={Physical Review D},
   publisher={American Physical Society (APS)},
   author={Williams, Michael J. and Veitch, John and Messenger, Chris},
   year={2021},
   month={ 05}
}

@article{Bilby,
   title={Bilby: A User-friendly Bayesian Inference Library for Gravitational-wave Astronomy},
   volume={241},
   ISSN={1538-4365},
   url={http://dx.doi.org/10.3847/1538-4365/ab06fc},
   DOI={10.3847/1538-4365/ab06fc},
   number={2},
   journal={The Astrophysical Journal Supplement Series},
   publisher={American Astronomical Society},
   author={Ashton, Gregory and Hübner, Moritz and Lasky, Paul D. and Talbot, Colm and Ackley, Kendall and Biscoveanu, Sylvia and Chu, Qi and Divakarla, Atul and Easter, Paul J. and Goncharov, Boris and et al.},
   year={2019},
   month={ 04},
   pages={27}
}

@misc{AvgNoise,
      title={Beyond Fisher: exact sampling distributions of the maximum-likelihood estimator in gravitational-wave parameter estimation}, 
      author={Michele Vallisneri},
      year={2011},
      eprint={1108.1158},
      archivePrefix={arXiv},
      primaryClass={gr-qc}
}

@article{GPULISA,
   title={GPU-accelerated massive black hole binary parameter estimation with LISA},
   volume={102},
   ISSN={2470-0029},
   url={http://dx.doi.org/10.1103/PhysRevD.102.023033},
   DOI={10.1103/physrevd.102.023033},
   number={2},
   journal={Physical Review D},
   publisher={American Physical Society (APS)},
   author={Katz, Michael L. and Marsat, Sylvain and Chua, Alvin J. K. and Babak, Stanislav and Larson, Shane L.},
   year={2020},
   month={ 07}
}


\begin{appendices}

\section{Minimising the Jeffreys' Divergence of Gaussians}\label{app:JeffDiv}

Let $\Theta$ be the space of parameters $\pvec$, with dimension $k=\mathrm{dim}(\Theta)$. Let $(\mathcal{D}_\f,\mathbb{C}_\f^{-1})$ be the data space and inner product operator pair for the \underline{f}ull dataset (subscript $_\f$). If the posterior defined on the parameter space, $p_\f(\pvec)$, is Gaussian, it can be written
\begin{equation}
p_\f(\pvec) = \left[(2\pi)^k|\fish_\f^{-1}|\right]^{-1/2}\exp\left[-\tfrac{1}{2}\boldsymbol{\Delta}^\T\fish_\f\boldsymbol{\Delta}\right] \, ,
\end{equation}
where $\fish_\f$ is the \gls{fim} of the full dataset, $|\fish_\f^{-1}|$ is the determinant of its inverse, $\boldsymbol{\Delta}\equiv \pvec-\hpvec$ and $\hpvec$ is the location of the peak of the posterior. Suppose there also exists the space $(\mathcal{D}_\s,\mathbf{M})$, with the posterior $p'_\s(\pvec)$, and where $\fish'_\s$ is the \gls{fim} in $(\mathcal{D}_\s,\mathbf{M})$, where $\mathbf{M}$ is some Ansatz matrix (we will set this equal to $\mathds{1}$ so that each sample is weighted equally). This is the (unmodified) downsampled subspace, that is: $\mathcal{D}_\s\subset\mathcal{D}_\f$.

Finally, let $(\mathcal{D}_\s, \mathbb{C}_\s^{-1}\!=\!m\mathbf{M})$ be a data space and inner product operator pair for a downsampled dataset, where $m$ is the constant we seek. Write the Gaussian posterior similarly on this pair, $p_\s(\pvec)$, as
\begin{equation}\label{subprobdef}
p_\s(\pvec) = \left[(2\pi)^k|\fish_\s^{-1}|\right]^{-1/2}\exp\left[-\tfrac{1}{2}\boldsymbol{\Delta}^\T\fish_\s\boldsymbol{\Delta}\right] \, ,
\end{equation}
where $\fish_\s$ is the \gls{fim} of the downsampled dataset.

Ideally, we would minimise the \gls{js} divergence, since this is a well-known, symmetric measure of distance between distributions and is impartial to either distribution, unlike the \gls{kl} divergence, which implies its first argument is the true distribution and its second is a sample distribution. However, we are choosing our posterior to be a replacement for the truth, not some sample distribution with minimal divergence from it, so we require a symmetric divergence measure. The \gls{js} divergence is defined in equation (\ref{JSdivdef}). It contains the integral of the log of a sum and only special cases are able to be integrated analytically. In order to avoid prioritisation of any given distribution, we shall attempt to minimise another symmetric divergence known as the \emph{Jeffreys'-divergence} instead, simply defined by
\begin{equation}\label{Jeffreys_div}
    D_{\mathrm{J}}\Big(p_\f\,\Big|\Big|\,p_\s\Big) \equiv D_{\mathrm{KL}}\Big(p_\f\,\Big|\Big|\,p_\s\Big) + (p_\s \leftrightarrow p_\f) \, .
\end{equation}
First however, we warm up by minimising the \gls{kl} divergence, before moving on to the Jeffreys' divergence.

We have the unknown $m$ in $(\mathcal{D}_\s, \mathbb{C}_\s^{-1})$, but $(\mathcal{D}_\s, \mathbf{M})$ and $(\mathcal{D}_\f,\mathbb{C}_\f^{-1})$ are known completely. Using the Gaussian approximations, we write down the \gls{kl} divergence to find the $m$ that minimises the \gls{kl} divergence between $p_\s(\pvec)$ and $p_\f(\pvec)$:
\begin{align*}
D_{\mathrm{KL}}(p_\f(\pvec)\,||\,p_\s(\pvec))\equiv&\int_\Theta \d^k\theta \, p_\f(\pvec)\ln\left(\frac{p_\f(\pvec)}{p_\s(\pvec)}\right) \\
=&\int_\Theta \d^k\theta \, p_\f(\pvec)\left[\ln\left(\sqrt{\frac{|{\fish_\s}^{-1}|}{|\fish_\f^{-1}|}}\right)+\tfrac{1}{2}\boldsymbol{\Delta}^\T(\fish_\s-\fish_\f)\boldsymbol{\Delta}\right] \\
=&\ln\left(m^{-k/2}\sqrt{\frac{|{\fish'}_\s^{-1}|}{|\fish_\f^{-1}|}}\right)+\tfrac{1}{2}\int_\Theta \d^k\theta \, p_\f(\pvec)\cdot\boldsymbol{\Delta}^\T(m\fish'_\s-\fish_\f)\boldsymbol{\Delta} \, ,\numberthis \label{approxapprox}
\end{align*}
since the integral of the posterior (which we take to be normalised) over parameter space is equal to 1. In the last line, we also used the definition of the \gls{fim} in equation (\ref{FIMdef}) and $\mathbb{C}_\s^{-1}\!=\!m\mathbf{M}$ to find that
\begin{equation}
F_{\s,ij}=h_{k,i}(\mathbb{C}_\s^{-1})_{kl}h_{l,j}=mh_{k,i}M_{kl}h_{l,j}=mF'_{\s,ij} \, ,
\end{equation}
or $\fish_\s = m\fish'_\s$, from which we obtain
\begin{equation}\label{fishdets}
|\fish_\s|=m^k|\fish'_\s| \, .
\end{equation}

The \gls{kl} divergence is a function of $m$, with the minimum being the $m$ at which the derivative with respect to $m$ vanishes (it is clear, given our restriction to considering Gaussians, that there will always be one extremum only, which is a minimum). The minimum is given by the $m=m_{\mathrm{KL}}$ solving $\frac{\d}{\d m}D_{\mathrm{KL}}(p_\f(\pvec)\,||\,p_\s(\pvec))=0$, which is
\begin{align*}\label{fact1}
m_{\mathrm{KL}} &= k\left( \int_\Theta \d^k\theta \, p_\f(\pvec)\cdot\boldsymbol{\Delta}^\T\fish'_\s\boldsymbol{\Delta} \right)^{-1} \, . \numberthis
\end{align*}
To compute the integral, let us change coordinates first by a translation $\theta_\mu\to{\theta}^\star_\mu$ that takes $\boldsymbol{\hat{\theta}}\to{\boldsymbol{\hat{\theta}}}^\star=\mathbf{0}$, centering the peak of the posterior at the origin, so that $\boldsymbol{\Delta}\to {\boldsymbol{\Delta}}^\star={\boldsymbol{\theta}}^\star$. Then perform a rotation ${\theta}^\star_\mu\to\widetilde{\theta}_\mu$ such that $\fish_\f^\star\to\widetilde{\fish}_\f$, where the matrix $\widetilde{\fish}_\f$ is diagonal, i.e., aligning the coordinate axes with the ellipsoid axes. We denote all objects in the final coordinates with a tilde accent. Both transformations are unitary resulting in a Jacobian determinant equal to 1, and if we further suppose that $\boldsymbol{\Theta}\sim \mathbb{R}^k$ then also $\widetilde{\boldsymbol{\Theta}}\sim \mathbb{R}^k$, so we can write the integral in (\ref{fact1}) as
\begin{align*}
\int_\Theta \d^k\theta \, p_\f(\pvec)\cdot\boldsymbol{\Delta}^\T\fish'_\s\boldsymbol{\Delta}&=\int_{\widetilde{\Theta}} \d^k\widetilde{\theta} \, p_\f(\widetilde{\pvec})\cdot\widetilde{\boldsymbol{\theta}}^\T\widetilde{\fish}'_\s\widetilde{\boldsymbol{\theta}}\\
&=\left[(2\pi)^k\left|\widetilde{\fish}_\f^{-1}\right|\right]^{-1/2}\int_{\widetilde{\Theta}} \d^k\widetilde{\theta} \, \exp\left[-\tfrac{1}{2}\widetilde{\boldsymbol{\theta}}^\T\widetilde{\fish}_\f\widetilde{\boldsymbol{\theta}}\right]\cdot\widetilde{\boldsymbol{\theta}}^\T\widetilde{\fish}'_\s\widetilde{\boldsymbol{\theta}}\\
&=\left[(2\pi)^k\left|\widetilde{\fish}_\f^{-1}\right|\right]^{-1/2}\int_{\widetilde{\Theta}} \d^k\widetilde{\theta} \, \exp\left[-\tfrac{1}{2}\sum_{l=1}^k\widetilde{F}_{\f,ll}\widetilde{\theta}_l^2\right]\sum_{i,j=1}^k\widetilde{\theta}_i\widetilde{F}'_{\s,ij}\widetilde{\theta}_j\\
&=\left[(2\pi)^k\left|\widetilde{\fish}_\f^{-1}\right|\right]^{-1/2}\sum_{i,j=1}^k\widetilde{F}'_{\s,ij}\int_{\widetilde{\Theta}} \d^k\widetilde{\theta} \, \prod_{l=1}^k\exp\left[-\tfrac{1}{2}\widetilde{F}_{\f,ll}\widetilde{\theta}_l^2\right]\widetilde{\theta}_i\widetilde{\theta}_j \, .
\end{align*}
Noting the well-known integral formulas:
\begin{equation*}
\int_{-\infty}^\infty \d x \;e^{-ax^2}=\sqrt{\pi/a}\, , \;\;\;\;\;\;\;\;\; \int_{-\infty}^\infty \d x \; x\,e^{-ax^2}=0\, , \;\;\;\;\;\;\;\;\; \int_{-\infty}^\infty \d x \; x^2\,e^{-ax^2}=\frac{\sqrt{2\pi}}{(2a)^{3/2}}\, , \numberthis \label{int_formulas}
\end{equation*}
we find that the only surviving terms in the $i,j$ sum are the $i=j$ terms. The integral becomes
\begin{align*}
\int_\Theta \d^k\theta \, p_\f(\pvec)\cdot\boldsymbol{\Delta}^\T\fish'_\s\boldsymbol{\Delta}&=\left[(2\pi)^k\left|\widetilde{\fish}_\f^{-1}\right|\right]^{-1/2}\sum_{i=1}^k\widetilde{F}'_{\s,ii}\int_{-\infty}^\infty \d\widetilde{\theta}_i \, \widetilde{\theta}_i^2\,e^{-\frac{1}{2}\widetilde{F}_{\f,ii}\widetilde{\theta}_i^2}\prod_{j\neq i}\int_{-\infty}^\infty \d\widetilde{\theta}_j \, e^{-\frac{1}{2}\widetilde{F}_{\f,jj}\widetilde{\theta}_j^2}\\
&=\left[(2\pi)^k\left|\widetilde{\fish}_\f^{-1}\right|\right]^{-1/2}\sum_{i=1}^k\widetilde{F}'_{\s,ii}\sqrt{\frac{2\pi}{\widetilde{F}^3_{\f,ii}}}\;\prod_{j\neq i}\sqrt{\frac{2\pi}{\widetilde{F}_{\f,jj}}}\\
&=\sum_{i=1}^k\frac{\widetilde{F}'_{\s,ii}}{\widetilde{F}_{\f,ii}} \label{int_pf_lnps1} \numberthis \, .
\end{align*}
If the matrix $\mathbf{G}$ diagonalises $\fish_\f$, that is, if $\fish_\f\to\widetilde{\fish}_\f=\mathbf{G}^{-1}{\fish}_\f\mathbf{G}$ for diagonal $\widetilde{\fish}_\f$, then the integral may be written
\begin{equation}\label{int_pf_lnps2}
\int_\Theta \d^k\theta \, p_\f(\pvec)\cdot\boldsymbol{\Delta}^\T\fish'_\s\boldsymbol{\Delta}=\sum_{i=1}^k\frac{G^{-1}_{ij}F'_{\s,jl}G_{li}}{G^{-1}_{ij}F_{\f,jl}G_{li}} \, ,
\end{equation}
or for general symmetric matrices $\mathbf{A}$ and $\mathbf{B}$, and with $\mathbf{P}$ diagonalising $\mathbf{A}$, we have
\begin{equation}\label{int_pf_lnps_gen}
\int_\Theta \d^k\theta \, e^{-\frac{1}{2}\boldsymbol{\Delta}^\T\mathbf{A}\boldsymbol{\Delta}}\cdot\boldsymbol{\Delta}^\T\mathbf{B}\boldsymbol{\Delta}=\sqrt{\frac{(2\pi)^k}{|\mathbf{A}|}}\sum_{i=1}^k\frac{P^{-1}_{ij}B_{jl}P_{li}}{P^{-1}_{ij}A_{jl}P_{li}} \, .
\end{equation}
Therefore our factor $m_{\mathrm{KL}}$, using (\ref{fact1}) is
\begin{align*}\label{fact2}
m_{\mathrm{KL}} &= k\left(\sum_{i=1}^k\frac{G^{-1}_{ij}F'_{\s,jl}G_{li}}{G^{-1}_{ij}F_{\f,jl}G_{li}} \right)^{-1}\, . \numberthis
\end{align*}
This in particular minimises $D_{\mathrm{KL}}(p_\f\,||\,p_\s)$, which should be noted is not equal to $D_{\mathrm{KL}}(p_\s\,||\,p_\f)$: the \gls{kl} divergence not symmetric. Let us now minimise the symmetric Jeffreys' divergence. With equation (\ref{Jeffreys_div}), this can be written as
\begin{align*}
D_{\mathrm{J}}(p_\f\,||\,p_\s)&\equiv\int_\Theta \d^k\theta \, (p_\f-p_\s)\cdot\ln\left(\frac{p_\f}{p_\s}\right) \\
&=\int_\Theta \d^k\theta \, (p_\f-p_\s)\left[\ln\left(\sqrt{\frac{|{\fish_\s}^{-1}|}{|\fish_\f^{-1}|}}\right)+\tfrac{1}{2}\boldsymbol{\Delta}^\T(\fish_\s-\fish_\f)\boldsymbol{\Delta}\right] \\
&=\tfrac{1}{2}\int_\Theta \d^k\theta \, (p_\f-p_\s)\cdot\boldsymbol{\Delta}^\T(m\fish'_\s-\fish_\f)\boldsymbol{\Delta} \, ,\numberthis
\end{align*}
again since the integral of both $p_\f$ and $p_\s$ over the whole space must be equal to 1. Writing out $p_\f$ and $p_\s$ in full we have
\begin{align*}
D_{\mathrm{J}}(p_\f\,||\,p_\s)&=\tfrac{1}{2}\int_\Theta \d^k\theta \, \left[ \sqrt{\frac{|\fish_\f|}{(2\pi)^k}}e^{-\frac{1}{2}\boldsymbol{\Delta}^\T\fish_\f\boldsymbol{\Delta}} - \sqrt{\frac{m^k|{\fish'}_\s|}{(2\pi)^k}}e^{-\frac{1}{2}m\boldsymbol{\Delta}^\T\fish'_\s\boldsymbol{\Delta}}\right] \cdot \boldsymbol{\Delta}^\T(m\fish'_\s-\fish_\f)\boldsymbol{\Delta} \\
&=C\int_\Theta \d^k\theta \, \left[ C'e^{-\frac{1}{2}\boldsymbol{\Delta}^\T\fish_\f\boldsymbol{\Delta}} - m^{k/2}e^{-\frac{1}{2}m\boldsymbol{\Delta}^\T\fish'_\s\boldsymbol{\Delta}}\right] \cdot \boldsymbol{\Delta}^\T(m\fish'_\s-\fish_\f)\boldsymbol{\Delta} \, ,
\end{align*}
where $C=\tfrac{1}{2}\sqrt{\frac{|{\fish'}_\s|}{(2\pi)^k}}$ and $C'=\sqrt{\frac{|\fish_\f|}{|{\fish'}_\s|}}$. Now differentiate with respect to $m$:
\begin{align*}
\frac{\d}{\d m} D_{\mathrm{J}}(p_\f\,||\,p_\s)&=C\int_\Theta \d^k\theta \, \left[ C'e^{-\frac{1}{2}\boldsymbol{\Delta}^\T\fish_\f\boldsymbol{\Delta}} - m^{k/2}e^{-\frac{1}{2}m\boldsymbol{\Delta}^\T\fish'_\s\boldsymbol{\Delta}}\right] \cdot \boldsymbol{\Delta}^\T\fish'_\s\boldsymbol{\Delta} \\
&\hspace{45pt} - \left[ \tfrac{k}{2} m^{k/2-1}e^{-\frac{1}{2}m\boldsymbol{\Delta}^\T\fish'_\s\boldsymbol{\Delta}} -\tfrac{1}{2}m^{k/2}\boldsymbol{\Delta}^\T\fish'_\s\boldsymbol{\Delta}e^{-\frac{1}{2}m\boldsymbol{\Delta}^\T\fish'_\s\boldsymbol{\Delta}}\right] \cdot \boldsymbol{\Delta}^\T(m\fish'_\s-\fish_\f)\boldsymbol{\Delta} \\
&=C\int_\Theta \d^k\theta \, C'e^{-\frac{1}{2}\boldsymbol{\Delta}^\T\fish_\f\boldsymbol{\Delta}} \cdot \boldsymbol{\Delta}^\T\fish'_\s\boldsymbol{\Delta} \\
&\hspace{45pt} - m^{k/2}e^{-\frac{1}{2}m\boldsymbol{\Delta}^\T\fish'_\s\boldsymbol{\Delta}}\left\lbrace\left[ 1 + \tfrac{k}{2} -\tfrac{1}{2}m\boldsymbol{\Delta}^\T\fish'_\s\boldsymbol{\Delta}\right] \cdot \boldsymbol{\Delta}^\T\fish'_\s\boldsymbol{\Delta} \right. \\
&\hspace{150pt} \left. - \left[ \tfrac{k}{2m} -\tfrac{1}{2}\boldsymbol{\Delta}^\T\fish'_\s\boldsymbol{\Delta}\right] \cdot \boldsymbol{\Delta}^\T\fish_\f\boldsymbol{\Delta} \right\rbrace  \\[4pt]
&=CC'\int_{{\Theta}} \d^k{\theta} \, e^{-\frac{1}{2}\boldsymbol{{\Delta}}^\T{\fish}_\f\boldsymbol{{\Delta}}} \cdot \boldsymbol{{\Delta}}^\T{\fish}'_\s\boldsymbol{{\Delta}} \\
&\hspace{45pt} - Cm^{k/2}\int_{{\Theta}} \d^k{\theta} \, e^{-\frac{1}{2}m\boldsymbol{{\Delta}}^\T{\fish}'_\s\boldsymbol{{\Delta}}} \cdot \boldsymbol{{\Delta}}^\T\left(\left[ 1 + \tfrac{k}{2}\right]{\fish}'_\s - \tfrac{k}{2m}{\fish}_\f\right)\boldsymbol{{\Delta}} \\
&\hspace{45pt} + \tfrac{1}{2}Cm^{k/2}\int_{{\Theta}} \d^k{\theta} \, e^{-\frac{1}{2}m\boldsymbol{{\Delta}}^\T{\fish}'_\s\boldsymbol{{\Delta}}}\left\lbrace m(\boldsymbol{{\Delta}}^\T{\fish}'_\s\boldsymbol{{\Delta}})^2 - \boldsymbol{{\Delta}}^\T{\fish}'_\s\boldsymbol{{\Delta}} \cdot \boldsymbol{{\Delta}}^\T{\fish}_\f\boldsymbol{{\Delta}} \right\rbrace \, . \numberthis \label{jeffderiv1}
\end{align*}
The first two integrals of the last line are computed easily using (\ref{int_pf_lnps_gen}) by making the relevant matrix substitutions;
\begin{equation}\label{int_1}
\int_{\bar{\Theta}} \d^k{\theta} \, e^{-\frac{1}{2}\boldsymbol{{\Delta}}^\T{\fish}_\f\boldsymbol{{\Delta}}} \cdot \boldsymbol{{\Delta}}^\T{\fish}'_\s\boldsymbol{{\Delta}} = \sqrt{\frac{(2\pi)^k}{|\fish_\f|}}\sum_{i=1}^k\frac{D^{-1}_{\f,ij}F'_{\s,jl}D_{\f,li}}{D^{-1}_{\f,ij}F_{\f,jl}D_{\f,li}}
\end{equation}
for diagonal $\mathbf{D}^{-1}_{\f}\fish_\f\mathbf{D}_\f$, and 
\begin{align*}
\int_{{\Theta}} \d^k{\theta} \, e^{-\frac{1}{2}m\boldsymbol{{\Delta}}^\T{\fish}'_\s\boldsymbol{{\Delta}}} \cdot \boldsymbol{{\Delta}}^\T & \left(\left[ 1 + \tfrac{k}{2}\right]{\fish}'_\s - \tfrac{k}{2m}{\fish}_\f\right)\boldsymbol{{\Delta}} \\[12pt]
&= \sqrt{\frac{(2\pi)^k}{|m\fish'_\s|}}\sum_{i=1}^k\frac{D^{-1}_{\s,ij}\left(\left[ 1 + \tfrac{k}{2}\right]F '_\s - \tfrac{k}{2m}F_\f\right)_{jl}D_{\s,li}}{D^{-1}_{\s,ij}mF'_{\s,jl}D_{\s,li}}\\
& = km^{-\frac{k+4}{2}}\sqrt{\frac{(2\pi)^k}{|\fish'_\s|}}\left(\left[ 1 + \tfrac{k}{2}\right]m - \tfrac{1}{2}\sum_{i=1}^k\frac{D^{-1}_{\s,ij}F_{\f,jl}D_{\s,li}}{D^{-1}_{\s,ij}F'_{\s,jl}D_{\s,li}}\right) \numberthis \label{int_2}
\end{align*}
for diagonal $\mathbf{D}^{-1}_{\s}\fish'_\s\mathbf{D}_\s$.

For the final integral in (\ref{jeffderiv1}) we again diagonalise the matrix in the exponent to make the integration simple. Similarly to how we transformed earlier, with a translation and a rotation, let us take $\theta_\mu\to\check{\theta}_\mu$ so that ${\fish}'_\s\to \check{\fish}'_\s$ where $\check{\fish}'_\s$ is diagonal, and such that $\boldsymbol{\check{\Delta}}=\boldsymbol{\check{\theta}}$. Then we can rewrite the integral as
\begin{align*}
I & = \int_{{\Theta}} \d^k{\theta} \, e^{-\frac{1}{2}m\boldsymbol{{\Delta}}^\T{\fish}'_\s\boldsymbol{{\Delta}}}\left\lbrace m(\boldsymbol{{\Delta}}^\T{\fish}'_\s\boldsymbol{{\Delta}})^2 - \boldsymbol{{\Delta}}^\T{\fish}'_\s\boldsymbol{{\Delta}} \cdot \boldsymbol{{\Delta}}^\T{\fish}_\f\boldsymbol{{\Delta}} \right\rbrace \\
& = \int_{\check{\Theta}} \d^k\check{\theta} \, e^{-\frac{1}{2}m\boldsymbol{\check{\theta}}^\T\check{\fish}'_\s\boldsymbol{\check{\theta}}}\left\lbrace m(\boldsymbol{\check{\theta}}^\T\check{\fish}'_\s\boldsymbol{\check{\theta}})^2 - \boldsymbol{\check{\theta}}^\T\check{\fish}'_\s\boldsymbol{\check{\theta}} \cdot \boldsymbol{\check{\theta}}^\T\check{\fish}_\f\boldsymbol{\check{\theta}} \right\rbrace \\
& = \int_{\check{\Theta}} \d^k\check{\theta} \, e^{-\frac{1}{2}m\boldsymbol{\check{\theta}}^\T\check{\fish}'_\s\boldsymbol{\check{\theta}}}\left\lbrace m\sum_{i=1}^k\check{F}'_{\s,ii}\check{\theta}_i^2\sum_{j=1}^k\check{F}'_{\s,jj}{\check{\theta}}_j^2 - \sum_{i=1}^k\check{F}'_{\s,ii}{\check{\theta}}_i^2 \sum_{j=1}^k\check{F}_{\f,jj}{\check{\theta}}_j^2 \right\rbrace \\
& = \int_{\check{\Theta}} \d^k\check{\theta} \, e^{-\frac{1}{2}m\boldsymbol{\check{\theta}}^\T\check{\fish}'_\s\boldsymbol{\check{\theta}}} \sum_{i=1}^k\check{F}'_{\s,ii}\check{\theta}_i^2\sum_{j=1}^k(m\check{F}'_{\s,jj} - \check{F}_{\f,jj}){\check{\theta}}_j^2 \\
& = \sum_{i=1}^k\check{F}'_{\s,ii}\sum_{j=1}^k(m\check{F}'_{\s,jj} - \check{F}_{\f,jj}) \int_{\check{\Theta}} \d^k\check{\theta} \, \prod_{l=1}^k e^{-\frac{1}{2}m \check{F}'_{\s,ll}\check{\theta}_l^2} \check{\theta}_i^2{\check{\theta}}_j^2 \\
& = \sum_{i=1}^k (m(\check{F}'_{\s,ii})^2 - \check{F}'_{\s,ii}\check{F}_{\f,ii}) \int_{-\infty}^\infty \d\check{\theta}_i \, e^{-\frac{1}{2}m \check{F}'_{\s,ii}\check{\theta}_i^2} \cdot \check{\theta}_i^4 \; \prod_{j\neq i} \int_{-\infty}^\infty \d^k\check{\theta} \, e^{-\frac{1}{2}m \check{F}'_{\s,jj}\check{\theta}_j^2} \\
&\hspace{20pt} + \sum_{i=1}^k\check{F}'_{\s,ii}\sum_{j\neq i}(m\check{F}'_{\s,jj} - \check{F}_{\f,jj}) \prod_{l\in\{i,j\}}\int_{-\infty}^\infty \d\check{\theta}_l \, e^{-\frac{1}{2}m \check{F}'_{\s,ll}\check{\theta}_l^2} \cdot \check{\theta}_l^2 \prod_{h\not\in\{i,j\}} \int_{-\infty}^\infty \d\check{\theta}_h \, e^{-\frac{1}{2}m \check{F}'_{\s,hh}\check{\theta}_h^2} \, ,
\end{align*}
where in the third line terms relating to the off-diagonal elements of $\check{\fish}_\f$ drop out since integrands of those terms are odd functions. In the last step, we split the sum into an $i=j$ part (first line) and a $i\neq j$ part (second line) for the $\check{\theta}_i^2{\check{\theta}}_j^2$ factor in the integrand. Along with (\ref{int_formulas}) we require the following well-known identity
\begin{equation}
\int_{-\infty}^\infty \d x \; x^4 e^{-ax^2} = \frac{3}{4}\sqrt{\frac{\pi}{a^5}} \, .
\end{equation}
Then we have that
\begin{align*}
I &= \sum_{i=1}^k (m(\check{F}'_{\s,ii})^2 - \check{F}'_{\s,ii}\check{F}_{\f,ii}) \frac{3}{4}\sqrt{\frac{\pi}{(\frac{1}{2}m \check{F}'_{\s,ii})^5}} \; \prod_{j\neq i} \sqrt{\frac{\pi}{\frac{1}{2}m \check{F}'_{\s,jj}}} \\
&\hspace{20pt} + \sum_{i=1}^k\check{F}'_{\s,ii}\sum_{j\neq i}(m\check{F}'_{\s,jj} - \check{F}_{\f,jj}) \prod_{l\in\{i,j\}} \sqrt{\frac{2\pi}{(m \check{F}'_{\s,ll})^3}} \prod_{h\not\in\{i,j\}} \sqrt{\frac{\pi}{\frac{1}{2}m \check{F}'_{\s,hh}}} \\[8pt]
&= 3m^{-\frac{k+4}{2}}\sum_{i=1}^k (m(\check{F}'_{\s,ii})^2 - \check{F}'_{\s,ii}\check{F}_{\f,ii}) (\check{F}'_{\s,ii})^{-2} \; \prod_{j=1}^k \sqrt{\frac{2\pi}{\check{F}'_{\s,jj}}} \\
&\hspace{20pt} + m^{-\frac{k+4}{2}}\sum_{i=1}^k\check{F}'_{\s,ii}\sum_{j\neq i}(m\check{F}'_{\s,jj} - \check{F}_{\f,jj}) \prod_{l\in\{i,j\}} (\check{F}'_{\s,ll})^{-1} \prod_{h=1}^k \sqrt{\frac{2\pi}{\check{F}'_{\s,hh}}} \\[8pt]
&= 3m^{-\frac{k+4}{2}}\sqrt{\frac{(2\pi)^k}{|\check{\fish}'_\s|}}\sum_{i=1}^k \left(m - \frac{\check{F}_{\f,ii}}{\check{F}'_{\s,ii}}\right) + m^{-\frac{k+4}{2}}\sqrt{\frac{(2\pi)^k}{|\check{\fish}'_\s|}}\sum_{i=1}^k\sum_{j\neq i}\left(m - \frac{\check{F}_{\f,jj}}{\check{F}'_{\s,jj}}\right) \\
&= m^{-\frac{k+4}{2}}\sqrt{\frac{(2\pi)^k}{|\check{\fish}'_\s|}}\left[ 2\sum_{i=1}^k \left(m - \frac{\check{F}_{\f,ii}}{\check{F}'_{\s,ii}}\right) + k\sum_{j=1}^k\left(m - \frac{\check{F}_{\f,jj}}{\check{F}'_{\s,jj}}\right) \right] \\
&= m^{-\frac{k+4}{2}}\sqrt{\frac{(2\pi)^k}{|\check{\fish}'_\s|}}\left[ (2+k) \left(km - \sum_{i=1}^k\frac{\check{F}_{\f,ii}}{\check{F}'_{\s,ii}}\right)\right] \, . \numberthis \label{I_int}
\end{align*}
Substituting (\ref{int_1}), (\ref{int_2}), and (\ref{I_int}) into (\ref{jeffderiv1}), we have
\begin{align*}
\frac{\d}{\d m} D_{\mathrm{J}}(p_\f\,||\,p_\s)&=CC'\sqrt{\frac{(2\pi)^k}{|\fish_\f|}}\sum_{i=1}^k\frac{D^{-1}_{\f,ij}F'_{\s,jl}D_{\f,li}}{D^{-1}_{\f,ij}F_{\f,jl}D_{\f,li}} \\
&\hspace{45pt} - Cm^{k/2}km^{-\frac{k+4}{2}}\sqrt{\frac{(2\pi)^k}{|\fish'_\s|}}\left(\left[ 1 + \tfrac{k}{2}\right]m - \tfrac{1}{2}\sum_{i=1}^k\frac{D^{-1}_{\s,ij}F_{\f,jl}D_{\s,li}}{D^{-1}_{\s,ij}F'_{\s,jl}D_{\s,li}}\right) \\
&\hspace{45pt} + \tfrac{1}{2}Cm^{k/2}m^{-\frac{k+4}{2}}\sqrt{\frac{(2\pi)^k}{|\check{\fish}'_\s|}}\left[ (2+k) \left(km - \sum_{i=1}^k\frac{\check{F}_{\f,ii}}{\check{F}'_{\s,ii}}\right)\right] \\
&=C\sqrt{\frac{(2\pi)^k}{|\fish'_\s|}}\left\lbrace\sum_{i=1}^k\frac{D^{-1}_{\f,ij}F'_{\s,jl}D_{\f,li}}{D^{-1}_{\f,ij}F_{\f,jl}D_{\f,li}} - km^{-2}\left(\left[ 1 + \tfrac{k}{2}\right]m - \tfrac{1}{2}\sum_{i=1}^k\frac{D^{-1}_{\s,ij}F_{\f,jl}D_{\s,li}}{D^{-1}_{\s,ij}F'_{\s,jl}D_{\s,li}}\right)\right. \\
&\hspace{80pt} \left.+ m^{-2}\left[ (1+\tfrac{k}{2}) \left(km - \sum_{i=1}^k\frac{\check{F}_{\f,ii}}{\check{F}'_{\s,ii}}\right)\right]\right\rbrace \\
&=\tfrac{1}{2}\left[\sum_{i=1}^k\frac{D^{-1}_{\f,ij}F'_{\s,jl}D_{\f,li}}{D^{-1}_{\f,ij}F_{\f,jl}D_{\f,li}} - m^{-2}\sum_{i=1}^k\frac{D^{-1}_{\s,ij}F_{\f,jl}D_{\s,li}}{D^{-1}_{\s,ij}F'_{\s,jl}D_{\s,li}}\right] \, .
\end{align*}
We want the $m=m_\mathrm{J}$ such that this derivative vanishes, so finally we see that the factor minimising the Jeffreys' divergence is given by:
\begin{equation}\label{app:Jeff_div_min}
m_\mathrm{J}= \sqrt{\sum_{i=1}^k\frac{D^{-1}_{\s,ij}F_{\f,jl}D_{\s,li}}{D^{-1}_{\s,ij}F'_{\s,jl}D_{\s,li}} \Bigg/ \sum_{i=1}^k\frac{D^{-1}_{\f,ij}F'_{\s,jl}D_{\f,li}}{D^{-1}_{\f,ij}F_{\f,jl}D_{\f,li}}} \; ,
\end{equation}
where recall that $\mathbf{D}_\s$ diagonalises $\fish'_\s$ and $\mathbf{D}_\f$ diagonalises $\fish_\f$. This is similar to $m_\mathrm{KL}$ in form but clearly balanced between the distributions.

\section{Exact Preservation of Fisher Information after Downsampling}\label{app:pres_fish}

\begin{multicols}{2}
We show here that the \gls{fim} can be preserved exactly after downsampling (if a non-pathological set of data points is selected) by either modifying the signal model or the sample noise covariance matrix (of remaining, pre-whitened samples). Pathological cases can occur in which a data point choice renders the precise recovery of proportionate Fisher information impossible. For example, if a certain signal contains more information on some parameter $\pvec_1$ overall, and the subset choice contains only samples that have more information on parameter $\pvec_2$, then no amount of signal/noise reweighting of those samples will be able to compensate for lost Fisher information on $\pvec_1$ without overcompensating for Fisher information on $\pvec_2$. This is an unlikely situation, but if necessary can be avoided in the initial sample selection stage. This method may allow retention of the qualitative features of the posterior in fewer samples than the single noise reweighting factor derived above.

Given an $\Nf$-dimensional data space and metric pair $(\Df,\icvm_\f)$ with \gls{fim} $F_{\f,ij}$, we suppose there exists an $\Ns$-dimensional data space and metric pair $(\Ds,\boldsymbol{\Gamma}^{-1}_\s)$ with \gls{fim} $F_{\s,ij}$, such that 
\begin{equation}\label{equalfims}
F_{\s,ij}=F_{\f,ij} \, .
\end{equation}
Then we have
\begin{align*}
{F}_{\s,ij}&={h}_{k,i}C^{-1}_{\f,kl}{h}_{l,j} \\
&=\bar{h'}_{k',i}\Gamma^{-1}_{\s,k'l'}\bar{h'}_{l',j} \, , \numberthis \label{fisher_gamma}
\end{align*}
where primes indicate objects on the $\Ns$-dimensional data subspace, the $\bar{h'}_{k',i}$ are the prewhitened model derivative vectors on $\Ds$, and $\boldsymbol{\Gamma}^{-1}_\s$ is the (to-be-determined) subspace metric in the whitened basis. Given a downsampling matrix $\mathbf{D}$, for example, as in equation (\ref{dsmat}), we could write
\begin{equation}
\bar{h'}_{k',i} = D_{k'm}C^{(-1/2)}_{\f,mk}h_{k,i} \, .
\end{equation}
The matrix $\xcvm_\s$ is unknown at this point, the form of which we wish to ascertain to satisfy equation (\ref{equalfims}). Thus, highlighting that one may modify the signal model itself similarly to the inner product operator on the subspace to achieve FIM preservation, let us further suppose that
\begin{equation}
{F}_{\s,ij}=(\omega \odot \bar{h}')_{k',i}\mathbf{Q}_{k'l'}(\omega \odot \bar{h}')_{l',j} \, ,
\end{equation}
where $\omega$ is a vector to be determined, and $\mathbf{Q}$ is some guess of sample covariance matrix. Since the samples are prewhitened, we may choose $\mathbf{Q}$ to be diagonal, and take as an ansatz $\mathbf{Q}=\mathds{1}$. With this ansatz, we can set the \gls{fim}s equal (however, note there may be some other $\mathbf{Q}$ that minimises the variation between the original and downsampled posterior over the entire parameter space). Proceeding with $\mathbf{Q}=\mathds{1}$, we have that
\begin{align*}
&(\omega \odot \bar{h}')_{k',i}\mathbf{Q}_{k'l'}(\omega \odot \bar{h}')_{l',j}\\
=\;&\sum_{k'=0}^{\Ns}\omega_{k'}^2 \bar{h}'_{k',i}\bar{h}'_{k',j}\\
=\;&(\bar{h}')^\T_{,i} \cdot \mathrm{diag}(\omega_1^2,\omega_2^2,...,\omega^2_{\Ns}) \cdot (\bar{h}')_{,j} \, ,\numberthis
\end{align*}
which we can compare with equation (\ref{fisher_gamma}) to see that
\begin{equation}\label{diagsubinvcovmat}
\xicvm_\s=\mathrm{diag}(\omega_1^2,\omega_2^2,...,\omega^2_{\Ns}) \, .
\end{equation}

If the \gls{fim} is an $n\times n$ matrix, i.e. there are $n$ parameters in our model, then since the \gls{fim} is symmetric, there will be $n_b=(n^2+n)/2$ unique equations in (\ref{equalfims}) given by combinations of the $i,j$ indices. We may define the weighting $\omega$ to be constrained to have the form of a sum of $n_b$ basis functions, for example:
\begin{equation}\label{weightfunc}
\omega^2_{k'} = \sum_{p=0}^{n_b-1}a_pt^p_{k'} \, ,
\end{equation}
where $t$ is the time, $t^p_{k'}$ is the ${k'}^\mathrm{th}$ time value raised to the $p^\mathrm{th}$ power, and the $a_p$ are constants. Thus we will have $n_b$ equations and $n_b$ unknowns (the $a_p$'s), and the set of $n_b$ simultaneous equations
\begin{widetext}
\begin{align*}
F_{\f,ij}&=\sum_{k'=0}^{\Ns}\left(\sum_{p=0}^{n_b-1}a_p t_{k'}^p \, \bar{h}'_{k',i}\bar{h}'_{k',j}\right)=\sum_{p=0}^{n_b-1}a_p\sum_{k'=0}^{\Ns} t_{k'}^p \, \bar{h}'_{k',i}\bar{h}'_{k',j} \, . \numberthis
\end{align*}
This can be written in matrix form $\mathbf{f}=\mathbf{Ba}$:
\begin{equation}\label{FIM_pres_mat}
\begin{pmatrix}
F_{\f,00} \\
\vdots \\
F_{\f,ij} \\
\vdots \\
F_{\f,nn} 
\end{pmatrix}=\begin{pmatrix}
(t^0\odot \bar{h}')^\T_{\;,0}\bar{h}'_{\;,0} & \hdots & (t^{p}\odot \bar{h}')^\T_{\;,0}\bar{h}'_{\;,0} & \hdots & (t^{n_b}\odot \bar{h}')^\T_{\;,0}\bar{h}'_{\;,0}\\
\vdots &  & \vdots & & \vdots\\
(t^0\odot \bar{h}')^\T_{\;,i}\bar{h}'_{\;,j} & \hdots & (t^{p}\odot \bar{h}')^\T_{\;,i}\bar{h}'_{\;,j} & \hdots & (t^{n_b}\odot \bar{h}')^\T_{\;,i}\bar{h}'_{\;,j}\\
\vdots &  & \vdots & & \vdots\\
(t^0\odot \bar{h}')^\T_{\;,n}\bar{h}'_{\;,n} & \hdots & (t^{p}\odot \bar{h}')^\T_{\;,n}\bar{h}'_{\;,n} & \hdots & (t^{n_b}\odot \bar{h}')^\T_{\;,n}\bar{h}'_{\;,n}
\end{pmatrix}\begin{pmatrix}
a_0 \\
\vdots \\
a_{p} \\
\vdots \\
a_{n_b}
\end{pmatrix} \, ,
\end{equation}
\end{widetext}
\noindent for $i\leq j\leq n \, ,$ and thereby, since we know the $F_{ij}$ and we can compute all the matrix elements, we can solve the above for the coefficients $a_p$, giving us the form of $\xicvm_\s$ using (\ref{diagsubinvcovmat}) and (\ref{weightfunc}). 
After computing $\xicvm_\s$, we define the downsampled log likelihood as
\begin{equation}\label{fim_pres_llhood}
    \ell = -\tfrac{1}{2} (\mathbf{D}\bar{\mathbf{r}})^\T\xicvm_\s(\mathbf{D}\bar{\mathbf{r}}) \, ,
\end{equation}
where again $\bar{\mathbf{r}}$ is the whitened residual and $\mathbf{D}$ is the sample selection matrix.

If we had selected as our ansatz some $\mathbf{Q}\neq\mathds{1}$, one would find that upon expanding, the resulting equations in the $a_p$ are quadric. These are difficult to solve analytically, and the difficulty of finding solutions quickly increases as the number of parameters of the model increases. Numerical methods could be employed in this case, to find solutions for the coefficients of the chosen basis function set in (\ref{weightfunc}).

The single noise reduction factor approach discussed in the previous section is equivalent to setting the \gls{fim}s approximately equal, with the assumption that $a_p\!\approx\!0$ for $p\!>\!0$ and finding the $a_0$ that minimises the variation in the \gls{fim}s. If a situation occurs in which this single factor is not precise enough to recover lost information, which could be expected when the Fisher information transmission rate changes significantly on timescales similar to the average duration between time samples\footnote{In other words, the signal cannot be said to be \emph{slowly evolving}. We can define \emph{Fisher information transmission rate} as the time derivative of the expectation value of the Fisher information: $\partial_t \langle F_{ij}(t) \rangle$.}, it is possible that the solution described here will be more robust (an interesting case to consider, for example, is extreme mass ratio inspirals in \gls{lisa}; the rate of transmission of Fisher information is far less uniform than slow evolution \gls{bhb}s, and approximate \gls{fim} recovery using a single factor may not be feasible).

In practice, in order to account for all $n_b$ FIM terms, the weight function $\omega_k^2$ (as we have decomposed it in terms of a polynomial in $t$) often contains negative values for any given random sample selection. Whilst this precisely replicates the FIM at the injected values, the negative weights are highly unphysical, and can adversely affect the likelihood at locations away from the signal parameters. For the purposes of our study, rather than work out a means by which to obtain strictly positive weights to reproduce the FIM (which might be as simple as using another basis) we shall just discard those random sample selections that yield negative weights $\omega_n^2<0$, and retry the random data point selections until only positive weights are found.

Furthermore, the matrix $\mathbf{B}$ is often ill-conditioned and inversion is unsafe. To ameliorate these problems, we transform to a basis such that the full dataset FIM is diagonal, then only preserve the diagonal entries, maintaining some accuracy whilst reducing some of these troublesome numerical issues.

\end{multicols}

\section{Numerical Marginalisation of Phase in Time-Domain}

For the latter problem, we can remove uninteresting (or \emph{nuisance}) parameters from the parameter space by marginalisation. The coalescence phase is one such parameter. The phase marginalised probability is given by
\begin{equation}
p'(\pvec'\,|\,\textbf{d})=\int_0^{2\pi} \d\phi_\mathrm{c}\, p(\pvec\,|\,\textbf{d})=\int_0^{2\pi} \d\phi_\mathrm{c}\,\frac{p(\pvec)\,p(\textbf{d}\,|\,\pvec)}{p(\mathbf{d})} \, ,
\end{equation}
where $\pvec'$ is the tuple of parameters excluding the coalescence phase, $\phi_\mathrm{c}$. If the prior on $\phi_\mathrm{c}$ is $(2\pi)^{-1}$, then the prior on $\pvec$ can be written
$$p(\pvec)=\frac{p'(\pvec')}{2\pi}\, ,$$
where $p'(\pvec')$ is the prior on $\pvec'$. Thus
\begin{equation}\label{marg_llhood}
p'(\pvec'\,|\,\textbf{d})=\frac{p'(\pvec')}{p(\mathbf{d})}\int_0^{2\pi} \d\phi_\mathrm{c}\,\frac{p(\textbf{d}\,|\,\pvec)}{2\pi} \, .
\end{equation}
Therefore, one can obtain the marginal probability simply by substituting a marginalised likelihood. Now since we can write the waveform in the form $h(t)=A(t)\cos(\phi(t)+\phi_\mathrm{c})$, then from equations (\ref{firstlikelihood}) and (\ref{orig_td_inner_prod}), we have that
\begin{align*}
    p(\textbf{d}\,|\,\pvec) &= \left[ (2\pi)^N\det(\cvm)\right]^{-1/2}\exp\left[-\tfrac{1}{2} \;\mathbf{\Big(d-h(\pvec)\Big)}^\T\icvm\mathbf{\Big(d-h(\pvec)\Big)}\right] \\
    &= \left[ (2\pi)^N\det(\cvm)\right]^{-1/2}\exp\left[-\tfrac{1}{2} \;\mathbf{\Big(d-\mathbf{A}\cos(\boldsymbol{\phi}'+\phi_\mathrm{c})\Big)}^\T\icvm\mathbf{\Big(d-\mathbf{A}\cos(\boldsymbol{\phi}'+\phi_\mathrm{c})}\right] \, ,
\end{align*}
where $\boldsymbol{\phi}'=\boldsymbol{\phi}'(\pvec')$. By trigonometric identities, we have
\begin{align*}
    \mathbf{\Big(d-h(\pvec)\Big)}^\T\icvm\mathbf{\Big(d-h(\pvec)\Big)} &= \mathbf{\Big(d-\mathbf{A}\cos(\boldsymbol{\phi}')\cos(\phi_\mathrm{c})+\mathbf{A}\sin(\boldsymbol{\phi}')\sin(\phi_\mathrm{c})\Big)}^\T \\ 
    &\hspace{20pt} \times \icvm \mathbf{\Big(d-\mathbf{A}\cos(\boldsymbol{\phi}')\cos(\phi_\mathrm{c})+\mathbf{A}\sin(\boldsymbol{\phi}')\sin(\phi_\mathrm{c})\Big)} \\
    &= A\cos^2(\phi_\mathrm{c})+ B \cos(\phi_\mathrm{c})\sin(\phi_\mathrm{c}) + C \cos(\phi_\mathrm{c})\\
    &\hspace{20pt}+D \sin^2(\phi_\mathrm{c}) + E \sin(\phi_\mathrm{c}) + F \, ,
\end{align*}
where
\begin{equation*}
\begin{aligned}[c]
    A &= \Big(\mathbf{A}\cos(\boldsymbol{\phi}')\Big)^\T \icvm \mathbf{A}\cos(\boldsymbol{\phi}')\\
    B &= -2\Big(\mathbf{A}\cos(\boldsymbol{\phi}')\Big)^\T \icvm \mathbf{\mathbf{A}\sin(\boldsymbol{\phi}')}\\
    C &= - 2\mathbf{d}^\T \icvm \mathbf{A}\cos(\boldsymbol{\phi}')
\end{aligned}
\hspace{60pt}
\begin{aligned}[c]
    D &= \mathbf{\Big(\mathbf{A}\sin(\boldsymbol{\phi}')\Big)}^\T \icvm \mathbf{A}\sin(\boldsymbol{\phi}')\\
    E &= 2\mathbf{d}^\T \icvm \mathbf{A}\sin(\boldsymbol{\phi}')\\
    F &= \mathbf{d}^\T \icvm \mathbf{d} \, .
\end{aligned}
\end{equation*}
Then the marginalised likelihood is
\begin{equation}
    \int_0^{2\pi} \d\phi_\mathrm{c}\,\frac{p(\textbf{d}\,|\,\pvec)}{2\pi} = \int_0^{2\pi} \d\phi_\mathrm{c}\,\frac{e^{-\frac{1}{2} \; \left( A\cos^2(\phi_\mathrm{c})+ B \cos(\phi_\mathrm{c})\sin(\phi_\mathrm{c}) + C \cos(\phi_\mathrm{c})+D \sin^2(\phi_\mathrm{c}) + E \sin(\phi_\mathrm{c}) + F \right)}}{2\pi \left[ (2\pi)^N\det(\cvm)\right]^{1/2}} \, .
\end{equation}

There does not appear to be a straightforward method to obtain a closed form expression of this integral, but a numerical integration can be achieved easily and performed very cheaply, as we shall show. The computationally expensive part is in evaluating $\boldsymbol{\phi}'$. There is a moderately expensive operation in computing inner products in the definition of the constants (given some $\pvec'$) $A$ to $F$. However,  notice that all the constants are easily computed by various inner products of the vectors $\cvm^{(-1/2)}\mathbf{d}$,
$\cvm^{(-1/2)} \mathbf{\mathbf{A}\sin(\boldsymbol{\phi}')}$, and 
$\cvm^{(-1/2)} \mathbf{A}\cos(\boldsymbol{\phi}')$, so only these three matrix operations are required, and the downsampling significantly reduces the cost of these matrix operations. For many likelihood evaluations, store the sines and cosines (and their squares) of a range of $\phi_\mathrm{c}$ values in memory so as not to reevaluate them on each likelihood function call.

One can now approximate the integral by a discrete sum. We tested numerically `pre-marginalised' posteriors against `post-marginalised' posteriors (that is, we acquired the full posterior from the non-marginalised model, and marginalised this) and found that the number of terms in the sum must be decreased to around 20 before the approximation begins to fail. The likelihood evaluation time only begins to increase noticeably (by a few percent) when the number of terms in the sum reaches around $10^4$. We used $10^3$ terms, giving an accurate marginalised likelihood with negligible likelihood evaluation time increase. The \gls{pe} convergence time, however, with the phase parameter removed, is significantly decreased by a factor of around three.

\end{appendices}

\end{document}